\documentclass[aps,twocolumn]{revtex4}
\usepackage{color}
\usepackage{psfrag}
\usepackage{dcolumn}
\usepackage{bm}
\usepackage{float}
\usepackage[latin1]{inputenc}
\usepackage[spanish,english]{babel}
\usepackage{amsfonts}
\usepackage{amssymb,epsf}
\usepackage{graphicx}
\usepackage{epstopdf}
\usepackage{amsmath,amssymb}
\usepackage{pdflscape}
\usepackage{adjustbox}
\usepackage{hyperref}
\usepackage{subcaption}
\usepackage[svgnames, dvipsnames]{xcolor}
\begin{document}
\title{Structure of Anisotropic Magnetized Neutron Stars in $f(\mathcal{R},T)$ Gravity with Realistic Equation of State}

\author{M. Savari\footnote{
        email address: m.savari@hafez.shirazu.ac.ir}, G. H. Bordbar\footnote{
        email address: ghbordbar@shirazu.ac.ir (corresponding author)}, J. Sedaghat\footnote{
        email address: j.sedaghat@shirazu.ac.ir }, A. Sheykhi\footnote{
        email address: asheykhi@shirazu.ac.ir }}

\affiliation{Department of Physics, College of Science, Shiraz
University, Shiraz 71454, Iran\\
Biruni Observatory, College of Science, Shiraz University, Shiraz
71454, Iran}

\begin{abstract}
In this study, within the framework of $f(\mathcal{R},T)$ modified
gravity, we investigate the influence of coupling parameter
($\lambda$), magnetic field  and anisotropy parameter on the
neutron star structure. This work employs an accurate equation of
state (EoS), derived from realistic microscopic calculations based
on the AV18 nucleon-nucleon potential, to compute the structure of
this compact object.
 Here, determination of Schwarzschild radius, compactness,  gravitational surface redshift
and Kretschmann scalar within the $f(\mathcal{R},T)$ gravity,
confirms that our theoretical results are consistent with the
observational constraints. While established physical EoSs within
the framework of Einstein gravity have successfully characterized
a broad range of compact objects, they remain inadequate in
explaining certain massive objects residing within the mass gap
($2.5-5 M_{\odot}$).  We show that some compact objects residing in the mass gap  interpreted as candidates of neutron stars within the framework of $f(\mathcal{R},T)$ gravity.  Motivated by the observation of objects
within this mass range, we employed a framework that
simultaneously incorporates  $f(\mathcal{R},T)=\mathcal{R}+2\lambda T$ gravity and
 Bowers-Liang anisotropy model.  Our study concludes that an increase in the star's
surface magnetic field (which dominates the softening of the EoS)
leads to a reduction in the maximum masses and  corresponding
radii. Furthermore, at a fixed $\lambda$, larger values of the
anisotropy parameter $\beta$ result in increased masses and radii.
Conversely, for fixed $\beta$ , a decrease in  $\lambda$ yields an
increasing trend in both masses and radii. Finally, we compare our
results with the observational data from LIGO/Virgo/KAGRA and
NICER, setting the parameters of the $f(\mathcal{R},T)$ theory and
anisotropy to successfully  reproduce the masses and radii of the GW170817, PSR J0952-0607 and PSR J0740+6620 
and the masses of the secondary components of GW190814 ($m_2=2.59^{+0.08}_{-0.09} M_{\odot} $) and GW200210-092254 ($m_2=2.83^{+0.47}_{-0.42}M_{\odot}$).


\end{abstract}

\maketitle
\section{Introduction}
Neutron stars (NSs) provide outstanding astrophysical laboratories
for investigating condensed matter under extreme conditions,
nuclear interactions, and high-energy particle phenomena
\cite{Steiner-1,Lattimer-2}. Their ultra-dense interior (composed
of degenerate neutron-rich matter) offers unique insights into the
interplay between quantum chromodynamics (QCD)
\cite{Reisenegger-5, Cenko-6}, general relativity (GR), and
magnetohydrodynamics (MHD) \cite{harding06-7}. Recent gravitational-wave observations have revealed compact objects with masses in the so-called mass gap region ($2.5-5 M_{\odot}$) \cite{abbott17-3,abbott19-4}. However, the nature of these objects remains uncertain, as they cannot be conclusively identified as either heavy NSs or small black holes.
 This leads to some
challenges, such as constructing the  EoS, which must account for
the extreme densities and strong interactions present in NS cores.
EoSs are generally formulated through some common approaches: (i) by adopting a polytropic model \cite{Horedt:2004,Pretel:2024, Sarmah:2024, Das:2023}, (ii)  piecewise-polytropic parameterization \cite {Read:2008iy, Huang:2024,Biswas:2023,Tsaloukidis:2023}, (iii)  explicitly incorporating physical interactions \cite{Dehman:2024,Burgio:2023, Lovato:2023, Martin:2022,Shang:2021}, (iv) spectral representations \cite{Lindblom:2010bb,Raaijmakers:2024,  Ecker:2023,Greif:2022,Somasundaram:2021}, and finally, the non-parametric methods using Gaussian processes \cite {Essick:2019ldf, Legred:2021hdx}.

 Although the
polytropic  approaches are simpler, they lacks a robust physical
description, and some of them are  particularly ill-suited for describing NS
interior \cite{drischler24-044320}.
 Consequently, we focus on EoS
derived from microscopic calculations utilizing accurate
nucleon-nucleon interaction potentials, such as AV18
\cite{bordbar98-714}.

While numerous physically motivated NS's EoSs exist \cite{Chabanat:1997un,Serot:1984ey,margueron18-025805}, they cannot support objects in the unknown mass gap region \cite{Ozel:2016oaf, Glendenning:2000, Baiotti:2016qnr, Lattimer:2012xj, Postnov:2014tza, Abbott:2020tfl} within the framework of Einstein gravity (Tolman-Oppenheimer-Volkoff (TOV) equations). 
It should be noted that for some objects falling into this mass gap region (such as the secondary components of GW190814  \cite{abbott19-4} and GW200210-092254  \cite{KAGRA:2021duu,Maurya_2024, Maurya:2025zds}) it remains unknown whether they are NSs or the smallest black holes. Our motivation is to demonstrate that such objects may be explainable as NSs within the framework of generalized 
 $f(\mathcal{R},T)$ gravity.
This outcome serves as the primary motivation for employing $f(\mathcal{R},T)$  gravity. Furthermore, we establish that the compact objects under investigation cannot be black holes, as their Schwarzschild radius is considerably smaller than their stellar radius,  and their  gravitational surface redshift remains below $0.4$. Hence, we investigate the structure of NSs within the framework of  $f(\mathcal{R},T)$ gravity.
 $f(\mathcal{R},T)$ gravity is a compelling modification of GR which has been proposed by Harko et al.\cite{harko11-8}. This theory extends GR by introducing a non-minimal coupling between geometry and matter \cite{bertolami08-9}, leading to novel dynamics absent in classical gravity. Unlike GR, where the field equations are derived purely from the metric, $f(\mathcal{R},T)$ gravity incorporates matter-dependent terms, offering a more flexible framework to address cosmic acceleration \cite{perlmutter98-10}, dark energy \cite{riess98-11}, and quantum-gravitational corrections \cite{birrell84-12}. The $f(\mathcal{R},T)$ Lagrangian generalizes Einstein's theory while preserving its geometric foundations, making it a more tractable alternative to other higher-order curvature theories \cite{nojiri17-13,capozziello20-14}.
Recent studies  \cite{shabani20-15, Carvalho:2021} have demonstrated the existence of exact solutions in $f(\mathcal{R},T)$ gravity, 
including anisotropic stars \cite{deb20-17} and accelerated expansion models \cite{singh16-18}, highlighting its potential to unify gravity with quantum field theory \cite{ zhao20-qg, lambiase21-qg, jamil12-qg} and dark sector effects \cite{barrientos22-19, moraes16-dark,zubair16-dark, shabani17-dark,yousaf17-dark}.

The presence of  magnetic field in NSs introduces pressure anisotropy, leading to distinct radial ($P_{r}$) and tangential ($P_{t}$) pressure components. However, directly measuring the internal magnetic field strength of magnetized NSs remains a significant challenge, necessitating the development of theoretical models to study the effects of such fields on stellar structure and physical parameters \cite{Cenko-6}.
%
 {Magnetized NSs are broadly classified into two categories based on their surface magnetic field strengths:(i) Pulsars, with surface magnetic fields ranging from $10^{12} G$ to $10^{14} G$  \cite{Rezzolla2018jee}. (ii) Magnetars, which exhibit even stronger surface fields, typically between $10^{14} G$ to $10^{16} G$ \cite{Rezzolla2018jee}. }
 This classification underscores the critical role of magnetic fields in shaping the observable and theoretical properties of NSs, motivating further research into their anisotropic behavior and extreme MHD regimes \cite{harding06-7}.

 In this paper, we investigate the equilibrium configurations of isotropic non-magnetized and anisotropic magnetized NSs within $f(\mathcal{R},T)$ gravity.
This article is organized as follows:
In Sec. \ref{Sec.II}, employing the lowest-order constrained variational (LOCV) method with the AV18 nuclear potential, we compute the EoS for neutron matter under two scenarios: (i) in the absence of a magnetic field and (ii) in the presence of a strong magnetic field. In Sec. \ref{Sec.III}, we present a brief overview of the formalism of $f(\mathcal{R},T)$ modified gravity, where we discuss the generalized field equations. Then, using the derived EoS, we solve the TOV equations \cite{Oppenheimer-56} within the framework of $f(\mathcal{R},T)$ gravity \cite{harko11-8} to determine the stellar properties of NS, including the maximum mass and corresponding radius.
In Sec. \ref{Sec.V}, we present the physical structure properties
of NSs such as Schwarzchild radius, Kretschmann scalar,
gravitational surface redshift, and in Sec. \ref{Sec.VI}, we
compare our theoretical predictions with observational data.
Finally, we present our concluding remarks in Sec. \ref{Sec.VII}.It is important to acknowledge that this analysis is contingent upon several specific modeling choices: the AV18  EoS  \cite{bordbar98-714}, the Bowers-Liang prescription for anisotropy  \cite{BowerLiang-48}, and a particular functional form of 
$f(\mathcal{R},T)$ modified gravity \cite{harko11-8}.  It is well recognized that predictions regarding the structure of compact objects are inherently model-dependent. Therefore, establishing robust constraints on the nature of compact objects within the mass gap would ideally require a systematic treatment of EoS uncertainties - for example, by marginalizing over a broad ensemble of EoS models and performing a joint statistical inference on both gravitational and matter-sector parameters. Nevertheless, the use of a precise and accurate EoS in the present work enhances the reliability of our results.
\section{ Neutron Star's Equation of State}\label{Sec.II}
The EoS plays a fundamental role in determining the structural properties of NSs, particularly when comparing isotropic non-magnetic configurations with anisotropic magnetized systems. In this study, we derive the EoS for both cases, first for isotropic, non-magnetized NS matter and second for anisotropic matter under different values of magnetic fields. To compute the energy of the system, we employ the lowest order constrained variational (LOCV) method, a well-established approach for interacting many-body systems.

At first, we consider the wave function of the interacting system as follows,
\begin{equation}
    \psi =F \phi, \label{01}
\end{equation}
where $\phi$ represents the non-interacting ground-state wave function, and $F$ is a correlation operator that incorporates two-body nuclear interactions,
\begin{equation}
    F =S\prod_{i>j}f(ij). \label{02}
\end{equation}
\begin{figure*}
    \includegraphics[width=9.11cm]{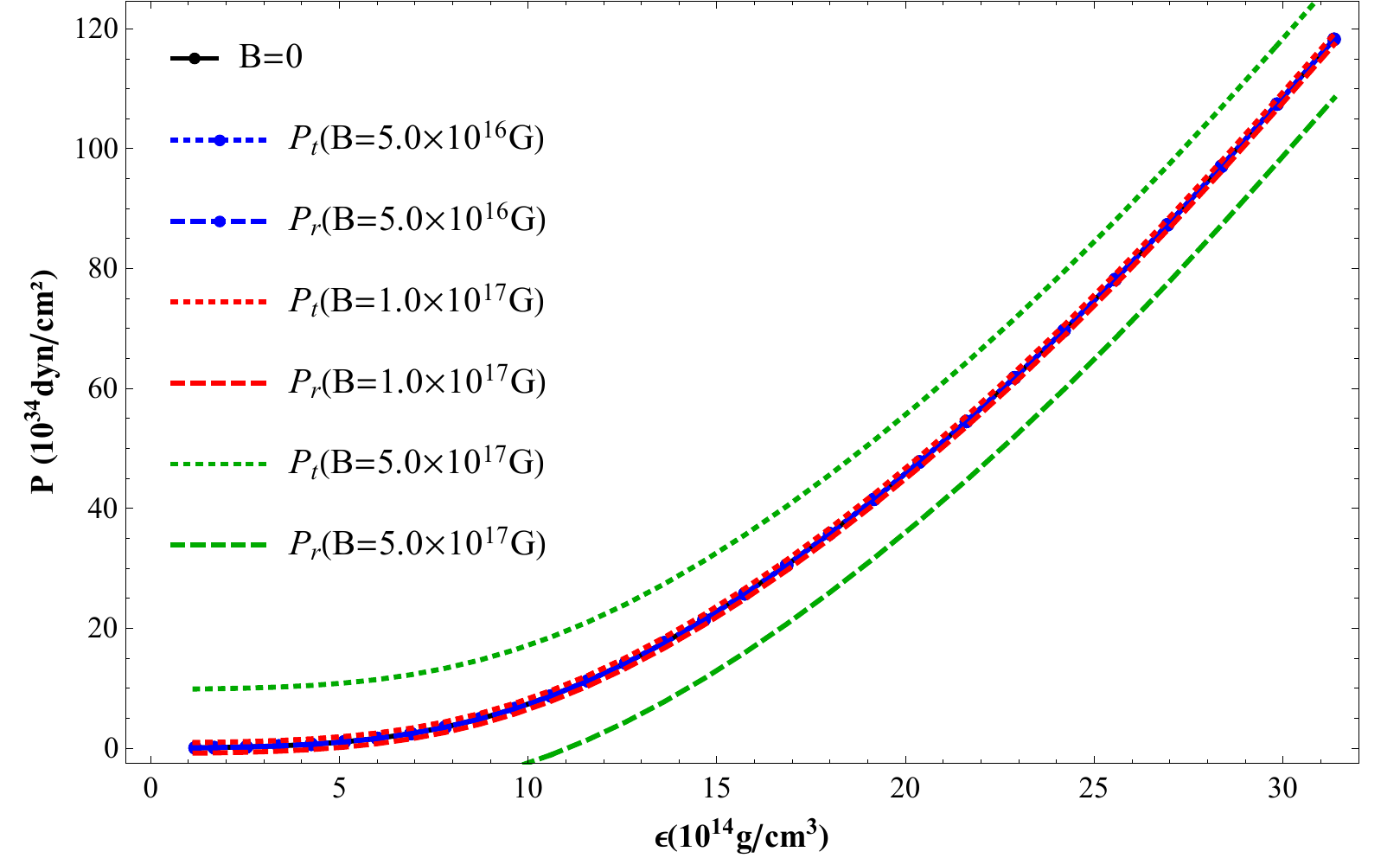}
    \includegraphics[width=8.71cm]{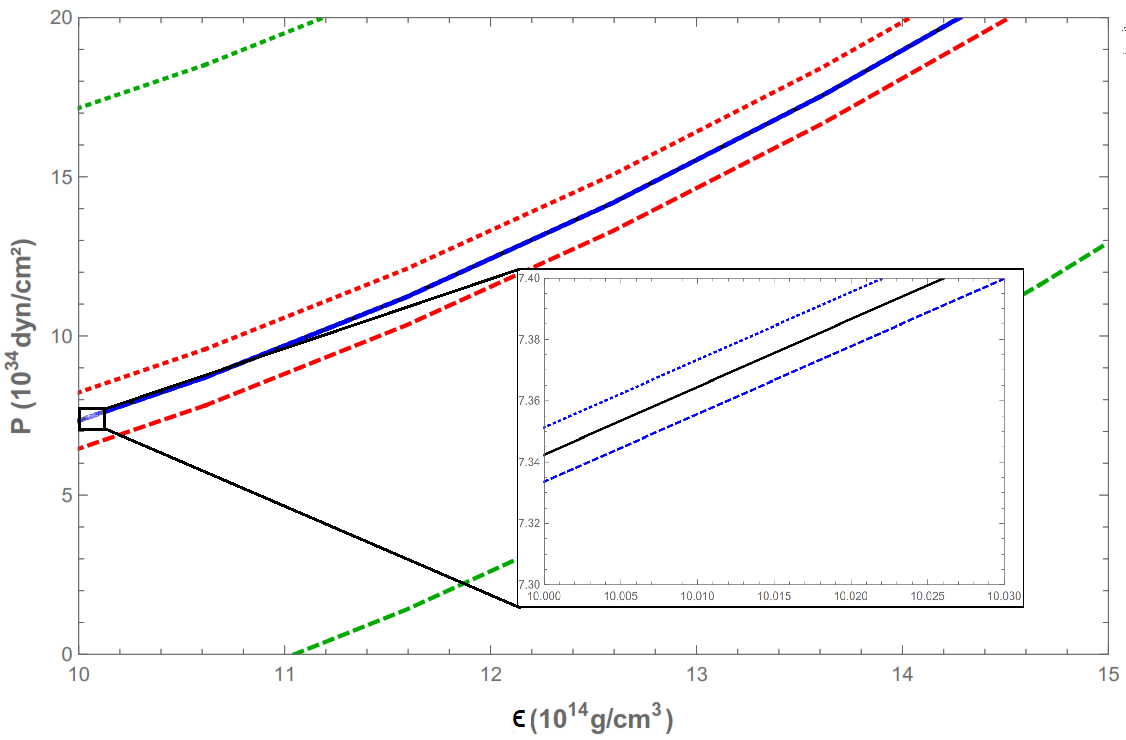}
    \caption{Left panel: The radial pressure and
tangential pressure vs energy density ($\epsilon = \frac{E}{V}$) of magnetized NS for three values of uniform magnetic fields. The pressure of non-magnetized case has been also given for comparison. The right plot is a magnification of the left one. }
    \label{101}
\end{figure*}
 Here $S$ and $f(ij)$ are symmetry operator and two-body correlation function, respectively.  This framework allows us to systematically account for both short-range repulsion and long-range attraction in neutron matter, providing a robust foundation for subsequent solutions of the structure equations \cite{wic78, car18}. To compute the expectation value of the energy, we employ a cluster expansion and retain contributions only up to the two-body energy terms, truncating higher-order correlations.
 This approximation captures the dominant one- and two-body energy contributions while ensuring computational tractability, a common approach in many-body systems where three-body and higher-order terms are expected to be subdominant,
\begin{equation}
    E([f])=\frac{\langle {\psi }|H|{\psi }\rangle }{N \langle {\psi }|{%
            \psi }\rangle }=E_{1}+E_{2}, \label{03}
\end{equation}
 where, $H=-\sum_{i} \frac{\hbar^{2}}{2m} \bigtriangledown
    _{i}^{2}+\sum_{i<j} V_{ij}$ denotes the Hamiltonian of the system and $E_{1}$ is the kinetic energy of the Fermi gas in the absence of magnetic field ($B=0$) which is calculated as follows,
\begin{equation}
    E_{1}=\sum_{i=+,-}\frac{3}{5}\frac{\hbar ^{2}k_{F}^{{(i)}^{2}}}{2m}\frac{%
        \rho ^{(i)}}{\rho },  \label{04}
\end{equation}%
where $k_{F}^{(i)}={(6\pi ^{2}\rho ^{(i)})}^{(\frac{1}{3})}$ is the Fermi
momentum of a neutron with spin projection $i$ and $\rho$ is the baryon number density. The two-body energy $E_{2}$ in the absence of magnetic field ($B=0$)  is also obtained as follows,
\begin{equation}
    E_{2}=\frac{1}{2N}\sum_{ij}\langle {ij}|\nu {(12)}|{ij-ji}\rangle , \label{05}
\end{equation}%
where
\begin{equation}
    \nu {(12)}=-\frac{\hbar ^{2}}{2m}[f(12),[\bigtriangledown
    _{12}^{2},f(12)]]+f(12)V(12)f(12).
    \label{06}
\end{equation}
In the above equation, $V(12)$ is the  two-body nucleon-nucleon interaction potential. We employ the accurate AV18 nuclear potential in all  calculations  \cite{wiringa95-38}. In  Ref. \cite{bordbar98-714}, the details of calculations for nuclear matter EoS have been presented.

In the presence of  strong magnetic fields, the total energy density of magnetized neutron matter must incorporate the contribution from the magnetic energy density, which modifies the EoS. Thus, the  total energy per particle is as follows,
\begin{equation}
    E(\rho ,B)=E_{1}+E_{2}-\mu _{n}B\delta,
    \label{07}
\end{equation}
where, $\mu _{n}=-9.662\times 10^{-24}  g.cm^2/s^2 G $ \cite{CODATA2018, Arnold:1947}, $B$ and $\delta \approx 0.1$ \cite{Liu:2021, Vidana:2002}  are the neutron magnetic moment, magnetic field, and spin polarization parameter, respectively.
To obtain the EoS of the system, we use the following equation,
\begin{equation}
    P(\rho ,B)=\rho ^{2}\biggl(\frac{\partial E(\rho ,B)}{\partial \rho }\biggr)%
  .  \label{08}
\end{equation}
In a strong magnetic field, the pressure of the NS differs in the radial and tangential directions, indicating that the NS undergoes the anisotropy. From the following equation, we can calculate  $P_{r}$ and   $P_{t}$, respectively \cite{Bordbar:2024yai-71},
%
\begin{equation}
    P_{t}=P_{r}+%
    \Delta,     \label{010}
\end{equation}
where $\Delta$ is the anisotropy parameter that will be discussed in the next sections. In order to examine the effect of  the density-dependent magnetic field (i.e $B(\rho)$), we here consider a Gaussian function of the  number density  as follows \cite{Bandyopadhyay1-51,Bandyopadhyay2-52},
\begin{equation}
    B(\rho)=B_{surf}+B_{0}\biggl[1-\exp \biggl(-{\eta (\frac{\rho }{\rho _{0}}%
        )^{\theta} \biggr)}\biggr], \label{011}
\end{equation}
where,  $\rho$ is normalized to the nuclear saturation density $\rho_{0} \approx 0.16  fm^{-3}$\cite{Horowitz:2020evx}. 
The magnetic field configuration is specified by its surface strengths $B_{\mathrm{surf}}=5.0\times10^{16} \mathrm{G}$ and $B_{\mathrm{surf}}=1.0\times10^{17}\mathrm{G}$, as well as its central value, $B_{0}=2.0\times10^{18}\mathrm{G}$. The spatial variation of the magnetic field is governed by two distinct decay regimes, defined by the parameters $\eta$ and $\theta$.
(i) Choosing $\eta=0.02$ and $\theta=3.0$ produces a rapid decrease from the core toward the surface.
(ii) In contrast, selecting $\eta=0.05$ and $\theta=2.0$ yields a more gradual attenuation, which we adopt as the slow-decay configuration throughout the remainder of this work (see Ref. \cite{Casali-53} for further details).
 Using this Gaussian model of magnetic field, we derive EoS for the system, incorporating the interplay between density-dependent magnetic effects and nuclear matter properties.
 \par
\begin{figure}[h!]
    \centering
    \includegraphics[width=8.1cm]{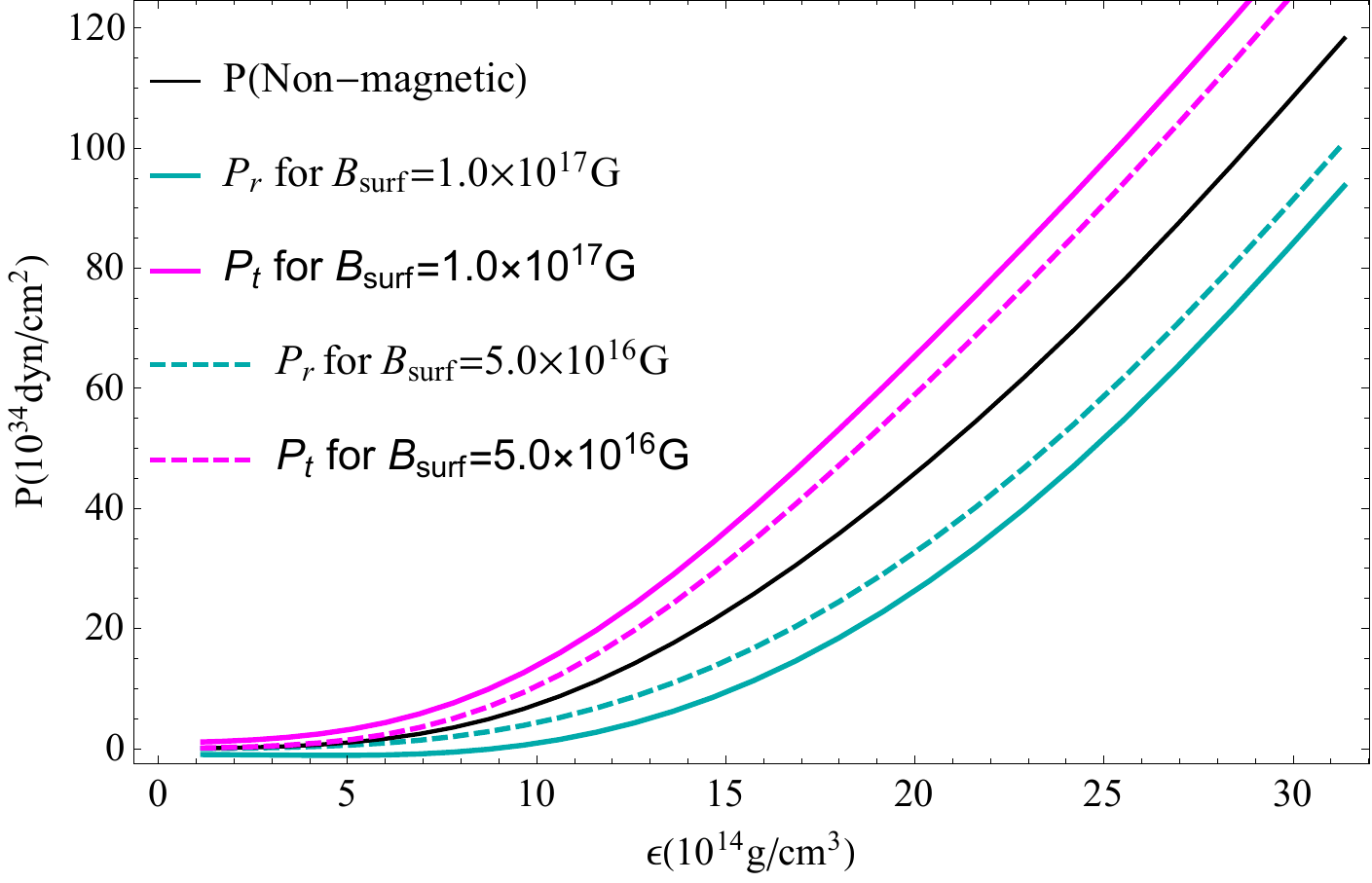}
    \caption{The radial pressure ($P_r$) and tangential pressure ($P_t$) vs energy density with the Gaussian form of magnetic field .  The non-magnetized case has been also given for comparison. }  \label{2}
\end{figure}

 Proceeding with our analysis, as it is seen from right panel of Fig. \ref{101}, the impact of magnetic fields below a certain value $10^{15}G$ on our calculations for the EoS is negligible.
 We classify them as 'weak magnetic fields', and fields exceeding $10^{15}G$ are termed 'strong magnetic fields'.
However, In the non-magnetized case ($B = 0$), the NS exhibits
isotropic pressure, with the radial  and tangential components
being identical ($P_{r} =P_{t}$), resulting in zero pressure
anisotropy ($\Delta \equiv P_{t}-P_{r}= 0$). For weakly magnetized
systems ($B<10^{15}G$), the pressure anisotropy $\Delta$ remains
negligible, as magnetic contributions to the energy-momentum
tensor are subdominant to nuclear interactions. However, in strong
magnetic fields ($B>10^{15}G$), the Lorentz force induces
significant anisotropy, manifesting in observable deviations in
the NS's bulk properties, such as its mass-radius relation,
compactness, surface redshift  and so on \cite{bocq9519,
frieb1219}. A clear indication of this phenomenon is shown in Fig.
\ref{101}. Notice that in Fig. \ref{101}, a uniform magnetic field
is used, whereas in Fig. \ref{2} a Gaussian form is employed.
It is obvious from these plots that as the  magnetic
field of the NS increases, the anisotropy in the magnetized NS increases. As the energy density
increases, there is a rising difference between the radial and tangential pressures in the Gaussian form of magnetic field. In Fig. \ref{2}, we have also compared the radial and tangential pressures with that of non-magnetized case.

\begin{figure}[h!]
    \centering
    \includegraphics[width=8.5cm]{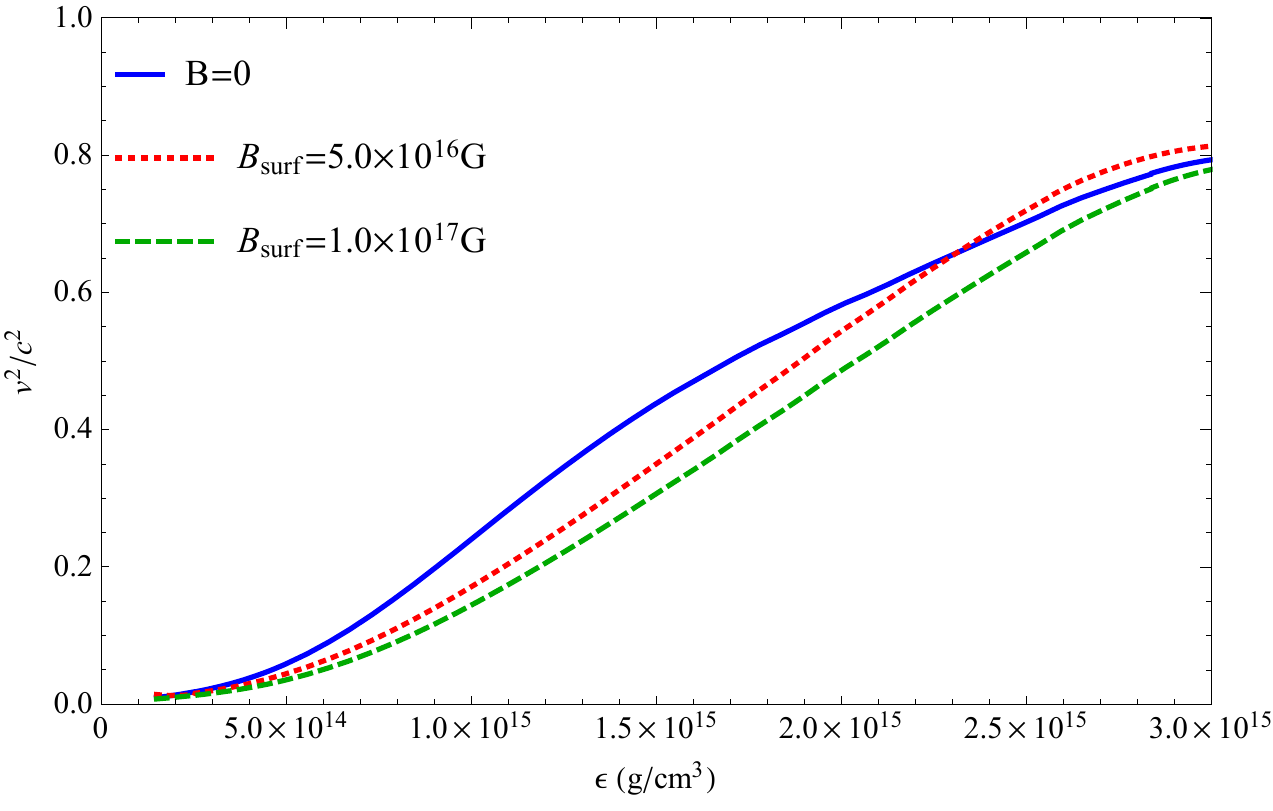}
    \caption{The ratio of the squared speed of sound to the squared speed of light in vacuum vs energy density for different values of magnetic fields within our EoS with Gaussian form of magnetic field. It is obvious that our EoS obey the causality conditions in all ranges of energy density.}  \label{1}
\end{figure}

To ensure the physical validity of our model, we examine the causality condition for both the isotropic and anisotropic magnetized cases, where the radial pressure component exhibits slow decay.
The condition of causality is given as follows,
\begin{equation}
    0\leqslant \frac{\nu ^{2}}{c^{2}}=\frac{1}{c^{2}}\biggl(\frac{dP_{r}(\rho ,B)}{d\rho }\biggr) \leqslant 1.  \label{012}
\end{equation}
In fact, for a physically admissible EoS, the speed of sound ($\nu$) must satisfy the condition $\nu^{2}<c^{2}$
 (where
$c$ is the speed of light in vacuum), ensuring that perturbations propagate causally within the NS matter.
 This constraint is crucial for ruling out superluminal energy-momentum transfer and maintaining thermodynamic consistency in the EoS (see Fig. \ref{1}). As demonstrated in the preceding analysis, the EoSs under consideration rigorously satisfy the causality condition. This ensures their physical validity and stability, thereby permitting their application in NS structure calculations.

  In the next section, we present the theoretical framework governing NSs as $f(\mathcal{R},T)$ gravity, including the derivation of the modified hydrostatic equilibrium equations.

\section{Formalism of $f(\mathcal{R},T)$ Modifeid Gravity }
 \label{Sec.III}
Following Harko et al.  \cite{harko11-8},
the action of the $f(\mathcal{R},T)$ modified theory of gravity is assumed as follows,
\begin{eqnarray}
S=\frac{1}{16\pi}\int
f\left(\mathcal{R},T\right)\sqrt{-g}\;d^{4}x+\int{\mathcal{L}_m\sqrt{-g}\;d^{4}x}\, \label{013}
\end{eqnarray}
where $f(\mathcal{R},T)$ is an arbitrary function of $\mathcal{R}$
and $T$ , the Ricci scalar and the trace of energy-momentum tensor
($T\equiv g^{\mu \nu }T_{\mu \nu }$), respectively.
$\mathcal{L}_m$ is the Lagrangian density of matter distribution,
which represents the possibility of non-minimal coupling between
matter and geometry. $g$ is the determinant of $g_{\mu \nu }$
metric. The energy-momentum tensor of matter defined as,
\begin{eqnarray}
T_{\mu \nu }=-\frac{2}{\sqrt{-g}}
\frac{\delta \left( \sqrt{-g}\mathcal{L}_m\right) }{\delta g^{\mu \nu }}.   \label{014}
\end{eqnarray}
It should be noted that in Eq. (\ref{014}) the derivative is taken
with respect to $g_{\mu \nu }$, not its derivatives. From the
variation of the action in Eq. (\ref{013}) with respect to the
metric, one gets the modified Einstein field equations in the
metric formalism \cite{harko11-8},
\begin{eqnarray}
&&f_{\mathcal{R}} \mathcal{R}_{\mu \nu } - \frac{1}{2}
f\left( \mathcal{R},T\right)  g_{\mu \nu }
+\left( g_{\mu \nu }\square -\nabla_{\mu }\nabla _{\nu }\right) f_{\mathcal{R}}
 \nonumber \\
&&  =8\pi G T_{\mu \nu}-f_{T} H_{\mu \nu},  \label{015}
\end{eqnarray}
where $f_{\mathcal{R}} =\partial
f\left(\mathcal{R},T\right) /\partial \mathcal{R}$ , $f_T =\partial
f\left(\mathcal{R},T\right) /\partial T$ and  $\square \equiv \nabla_{\alpha }\nabla^{\alpha }$ is the d'Alembert operator with $\nabla_{\alpha }$ representing the covariant derivative, and $H_{\mu \nu}\equiv T_{\mu \nu }+\Theta _{\mu \nu}$. Here $\Theta _{\mu \nu}$ tensor is defined as,
\begin{eqnarray}
\Theta_{\mu \nu}\equiv g^{\alpha \beta }\frac{\delta T_{\alpha \beta
}}{\delta g^{\mu \nu}}\,=-2T_{\mu \nu}+g_{\mu \nu }\mathcal{L}_m-2g^{\alpha \beta }
\frac{\partial ^2\mathcal{L}_m}{\partial g^{\mu \nu }\partial g^{\alpha \beta
}}\,.   \label{016}
\end{eqnarray}
Using the covariant derivative of Eq. (\ref{015}) (we mean the
Bianchi identity), one can obtain the following modified equation
for the four-divergence of the energy-momentum tensor
\cite{barrientos14-63},
\begin{eqnarray}
\nabla ^{\mu }T_{\mu \nu }=\frac{f_{T} }{8\pi  -f_{T} }
\left[ H_{\mu \nu} \nabla^{\mu }
\ln f_{T}\,
-\frac{1}{2}\nabla_{\nu} T+\nabla ^{\mu }\Theta _{\mu \nu }\right ].    \label{017}
\end{eqnarray}
This equation states that in $f(\mathcal{R},T)$ theory of gravity, $T_{\mu \nu }$ is not covariantly conserved. In fact, there is an extra force that makes the test particles deviate from the GR geodesics, and this is hugely important, because we do not need Dark Energy in the universe any more.
Furthermore, the trace of Eq. (\ref{015}) leads to,
\begin{eqnarray}
&&f_{\mathcal{R}} R+3\square f_{R} -2
f\left( \mathcal{R},T\right)
=8\pi G T-f_{T}
H\, , \
\label{018}
\end{eqnarray}
where $H\equiv g^{\mu \nu}H_{\mu \nu}$. In present study, because
there is no unique definition of the Lagrangian density of the
matter distribution, we simply assume that $\mathcal{L}_m=+P$ \cite{Pappas:2022gtt-64};
where $P$ stands for the isotropic pressure, then we can write
$\Theta _{\mu \nu}=-2T_{\mu \nu}+P g_{\mu \nu }$. As
you see from Eq. (\ref{017}) the conservation of $T_{\mu \nu}$ is
restored in the limit of $f(\mathcal{R})$ gravity i.e. when we put
$f_{T}=0$. In principle, for an arbitrary function of
$f(\mathcal{R},T)$ model, the components of energy-momentum tensor
will be too complicated, and consequently obtaining analytic
solutions will be impossible. In our work we choose specific
models as $f(\mathcal{R},T)=\mathcal{R}+h(T)$ which are linear in
Ricci scalar and do not involving mixing terms between
$\mathcal{R}$ and $T$. However the field equations (from Eqs.
(\ref{015}) and  (\ref{018})) will be as follows,
\begin{eqnarray}
G_{\mu
\nu}=\frac{8\pi
G}{c^4}T_{\mu\nu} +\frac{h}{2}g_{\mu \nu}+h_{T}(T_{\mu \nu }- P g_{\mu \nu }) ,   \label{19}
\end{eqnarray}
where $h_T =\partial h(T) /\partial T$. In the present paper, to
avoid the complicated field equations which may hinder the search
for exact analytic solutions, we choose the famous model in the
literature which is \cite{Pappas:2022gtt-64}:
\begin{eqnarray}
f(\mathcal{R},T)=\mathcal{R}+2\lambda T,    \label{20}
\end{eqnarray}
where $\lambda$ is the dimensionless constant free parameter,
which is the non-minimal coupling parameter of matter distribution
and curvature. Notice that for non-linear models of
$f(\mathcal{R},T)$ theory, one may in principle, deal with the
overly complicated field equations \cite{Pappas:2022gtt-64}. In
the linear regime of  $f(\mathcal{R},T)$ theory we choose specific
values of the  free parameter $\lambda$ , Our purpose is to
establish a model to fit with latest observations of
LIGO/Virgo/KAGRA\cite{abbott17-65,abbott18-66} and NICER
\cite{riley19-67}.

In the next subsection we will expand the  isotropic TOV equations to study the inner structures of the  non-magnetized NSs.

\subsection{Modified Isotropic TOV Equations in $f(\mathcal{R},T)$ Gravity} \label{SubSec.IIIA}
\begin{figure}[h!]
    \centering
    \includegraphics[width=8.5cm]{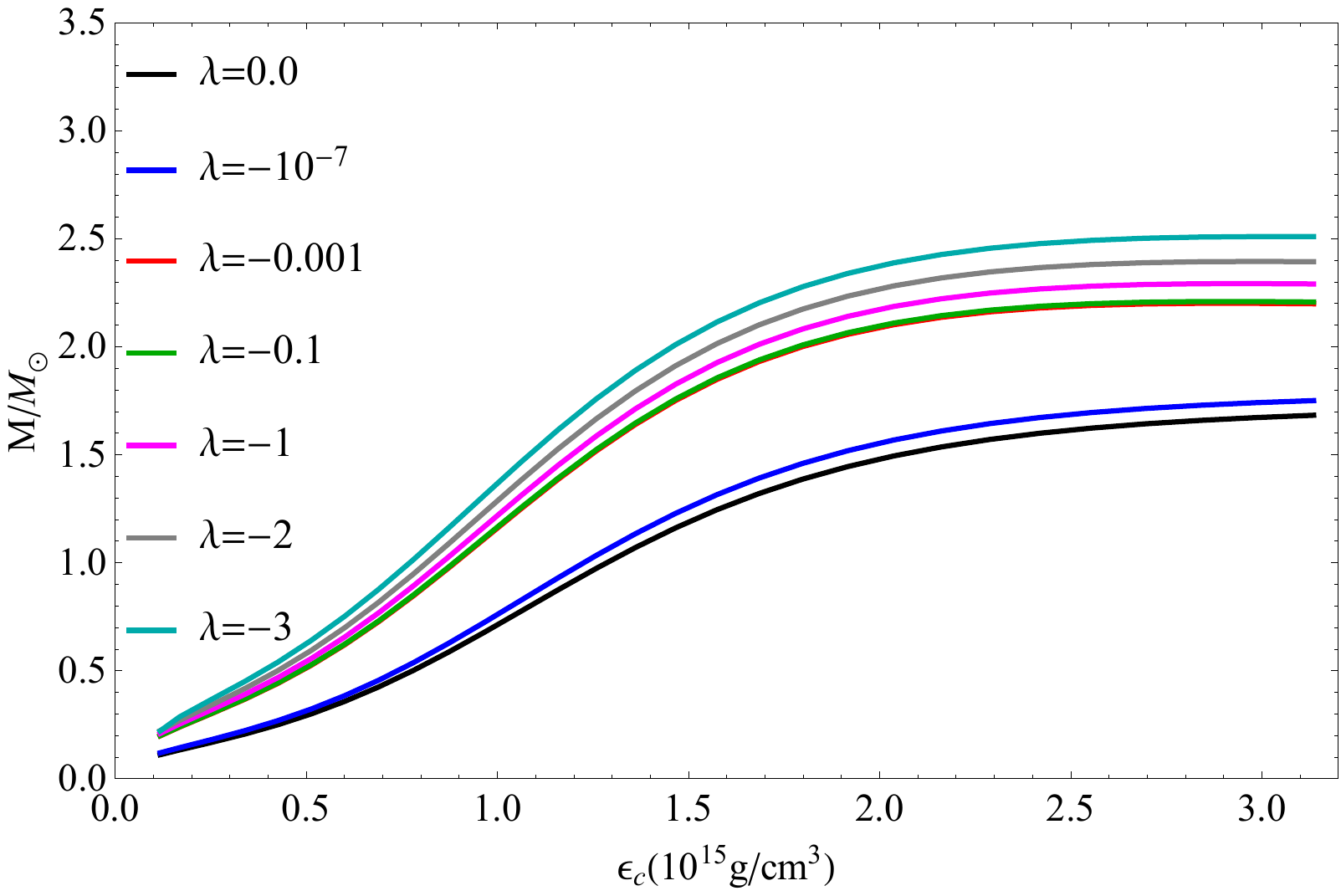}
    \caption{ The mass-central energy density relation to extract  Fig.\ref{6} in an  isotropic, non-magnetic NS.  Note that for $\lambda=0$ (The Black curve) the NS's mass reproduces GR values like: $1.682M_{\odot}$.}  \label{7}
\end{figure}

\begin{figure}[h!]
    \centering
    \includegraphics[width=8.5cm]{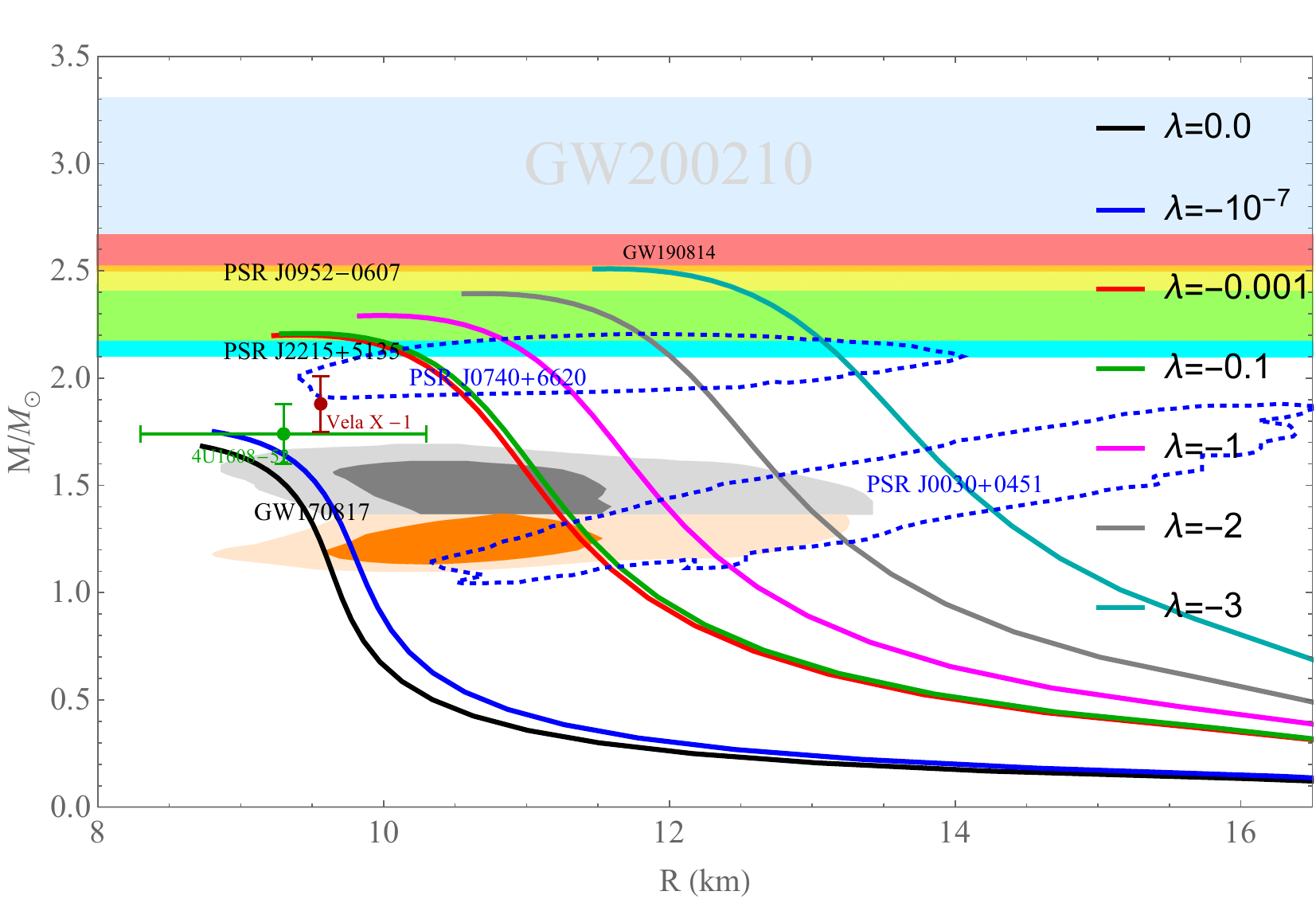}
    \caption{ The mass-radius relations for  isotropic non-magnetized NS, Notice that for  $\lambda =0$, the theory reduced to the non-magnetized GR (The black curve).  The mass of {pulsar PSR J0952-0607 (yellow), mass of pulsar PSR J2215+5135 } (cyan), and their overlapping region (green). The central hatched area corresponds to the mass and radius of the components of GW170817.}   \label{6}
\end{figure}

We seek spherically symmetric solutions of the $f(\mathcal{R},T)$ field equations in the interior of a non-rotating NS, so we consider the most general static and spherically symmetric line element as follows,
\begin{eqnarray}
ds^2=-e^{\Phi (r)}dt^2+e^{\Psi (r)} dr^{2}+r^{2}d{\Omega}_{2}^{2},  \label{21}
\end{eqnarray}
where $\Phi (r)$ and $\Psi (r)$ are functions of the radial
coordinates only (also called the potentials of the metric) and
$d{\Omega}_{2}^{2}$ is the surface element of a 2-sphere. Here,
the signature of the metric is $(-,+,+,+)$. Substituting the
perfect-fluid $T_{\mu \nu }$ from Eq. (\ref{28}) and the above
line element into Eq. (\ref{19}) yields the system of field
equations as,
\begin{eqnarray}
&&G^{0}_{0}=\frac{1}{r^2}\frac{d}{dr}(re^{-\Psi}) - \frac{1}{r^2}
 \nonumber \\
 &&= 8\pi G\epsilon c^2+\lambda (3\epsilon c^2- P)\equiv  \frac{8\pi G}{c^2}\epsilon_{eff}\, ,     \label{22}
\end{eqnarray}
\begin{eqnarray}
&&G^{1}_{1}=e^{-\Psi}\left( \frac{1}{r}\Phi' + \frac{1}{r^2} \right)- \frac{1}{r^2}
 \nonumber \\
 &&= \frac{8\pi G}{c^4} P +\lambda ( 3P -\epsilon c^2)\equiv  \frac{8\pi G}{c^4} P_{eff}.     \label{23}
\end{eqnarray}
As it can be seen from the form of  $\epsilon_{eff}$ (Eq. (\ref{22})) and $P_{eff}$ (Eq. (\ref{23})), the overall effect of modified gravity on the system is to introduce a mixing between the $\epsilon$ and $P$ of matter in the right hand side of Eqs. (\ref{22}) and (\ref{23}). Hence, according to Ref. \cite{Pappas:2022gtt-64} and Eq. (\ref{017}) we obtain the following relation,
\begin{eqnarray}
P'+(\epsilon c^2 + P)\frac{1}{2}\Phi' =\frac{\lambda}{8\pi +2\lambda}({\epsilon}'c^2-P') ,  \label{24}
\end{eqnarray}
which obviously reduces to GR's conservation equation; when $\lambda=0$. Notice that the prime denotes derivations in respect of $r$ (the radial coordinate). Upon introducing the mass function,
\begin{eqnarray}
e^{-\Psi (r)}\equiv 1-\frac{2Gm(r)}{rc^2},  \label{25}
\end{eqnarray}
the $00$ component of the field equations yields the mass of the star enclosed within a radius $r$ as,
\begin{eqnarray}
&&m(r)=4\pi \int_{0}^{r}\tilde{r}^{2}\epsilon_{eff}c^2(\tilde{r})d\tilde{r}
 \nonumber \\
 &&=  4\pi   \int_{0}^{r}\tilde{r}^{2}\epsilon(\tilde{r})c^2d\tilde{r}-\frac{\lambda}{2} \int_{0}^{r}\tilde{r}^{2}(P(\tilde{r}) -3\epsilon(\tilde{r})c^2)d\tilde{r}
 \nonumber \\
&&=m^{GR}(r)+m^{f(\mathcal{R},T)}(r)
\label{26}
\end{eqnarray}
In GR, $m(r)$ is determined by $\epsilon$ (the first term $m^{GR}(r)$), but in linear $f(\mathcal{R},T)$ gravity $P$ and $\rho$ contribute to the total mass (the second term $m^{f(\mathcal{R},T)}(r)$). To satisfy the strong energy condition($3\epsilon c^2- P\leqslant 0$)  and obtain mass values exceeding the GR predictions, the parameter $\lambda$ must be assigned negative values. In simple terms, The coupling of pressure in the mass equation results in the enhancement of the maximum mass for negative values of the parameter $\lambda$. Motivated by this effect, we adopt exclusively negative values for $\lambda$ in this study. Our focus is thereby directed toward the regime of super GR masses, deliberately setting aside the investigation of masses lower than the general relativistic predictions (see Figs.\ref{7} and \ref{6}). Therefore the differential form of Eq. (\ref{26}) is,
\begin{eqnarray}
\frac{dm(r)}{dr}=4\pi r^{2} \epsilon (r) c^2+\frac{\lambda}{2}(3\epsilon c^2- P)r^{2}.     \label{27}
\end{eqnarray}
Obviously if $\lambda$ tends to zero, the second term of the above equation vanishes, that means it will reduces to GR. The boundary conditions for the stellar structure equations are specified as follows; (i) at the origin $(r=0)$, the enclosed mass vanishes $(m(0)=0)$, ensuring regularity at the core; and (ii) at the surface $(r=R)$, the mass function converges to the total gravitational mass of the configuration $(m(R)=M)$, where
$R$ denotes the NS's radius. These conditions guarantee a physically consistent solution, matching the interior spacetime to an exterior Schwarzschild metric in the vacuum beyond NS's radius.
\begin{table}[h!]
    \caption{{\protect\small { Maximum masses and corresponding radii of isotropic non-magnetized NS for different coupling parameters ($\lambda$) in $f(R,T)$ gravity.}}}
    \label{results1}\centering
    \par
    \begin{adjustbox}{width=.32\textwidth}
        \begin{tabular}{|c|c|c|c|c|c|c|}
            \hline
            $ \lambda $ & ${M_{max}}\ [M_{\odot}]$ & $R[km]$ \\ \hline
            $0.0$ & 1.682 & 8.731  \\ \hline
            $-10^{-7}$ &1.750 & 8.814  \\ \hline
            $-0.001$ & 2.197 & 9.577  \\ \hline
            $-0.1$ & 2.208& 9.514 \\ \hline
         $-1.0$ & 2.291& 9.947 \\ \hline
            $-2.0$ & 2.394 & 10.684  \\ \hline
            $-3.0$ & 2.509 & 11.606  \\ \hline

        \end{tabular}
    \end{adjustbox}
\end{table}

We model the stellar configuration as a perfect fluid in hydrostatic equilibrium described by an energy-momentum tensor  of the form,
\begin{eqnarray}
T_{\mu \nu }=(\epsilon c^2 + P) u_{\mu}u_{\nu}+  P g_{\mu \nu},     \label{28}
\end{eqnarray}
 where the energy density
$ \epsilon $ and pressure  $P$ are static (time-independent). For a spherically symmetric, static spacetime, the fluid's four-velocity reduces to $u_{\mu}=(e^{-\Phi /2},0,0,0)$ , satisfying the normalization condition $u_{\mu}u^{\mu}=-1$ . Consequently, the only non-vanishing components of the  energy-momentum tensor  are the diagonal terms, reflecting the absence of shear stresses or energy flux in the equilibrium configuration as follows,
\begin{eqnarray}
T_{0}^{0}=-\epsilon c^2, \ T_{j}^{i}= P \delta_{j}^{i} .   \label{29}
\end{eqnarray}

By combining the field equations given in Eqs. (\ref{22}) and (\ref{23}) with the hydrostatic equilibrium condition  Eq. (\ref{24}), we derive the modified  TOV  equation within the framework of linear $f(R,T)$ gravity \cite{Pappas:2022gtt-64},
\begin{eqnarray}
\frac{dP}{dr}&=&-\frac{(\epsilon c^2+P)}{(1+\zeta)}\frac{\left[\frac{G}{c^2}m(r)+4\pi\,P \,r^3 - \frac{1}{2}\lambda \,r^3\,(\epsilon c^2-3P)\right]}{r\left[r-\frac{2Gm(r)}{c^{2}}\right]}\nonumber\\
&&+ \frac{\zeta c^2}{(1+\zeta)}\frac{d\epsilon }{dr}\,,  \label{30}
\end{eqnarray}
where $\zeta \equiv 8\pi +2\lambda$. It should be noted that for the case where $\lambda=0$, Eq. (\ref{30}) reduces to the standard TOV equation in GR. Eqs. (\ref{27}) and (\ref{30}) are the set of equations that we call them "modified isotropic TOV equations in $f(\mathcal{R},T)$
gravity". To determine the structure of a relativistic spherically symmetric NS in hydrostatic equilibrium, the TOV equations must be solved subject to the boundary conditions:(i) Surface condition: The pressure vanishes at the NS's surface radius, $P(R)=0$. (ii) Central condition: The pressure attains its maximum value at the center, $P(0)=P_{c}$ where $P_{c}$ denotes the central pressure. By numerically integrating the TOV equations under these boundary conditions, one obtains the structure of NS.

Our results are given in Figs. \ref{7} and \ref{6}. As it is seen, when the central energy density increase, the mass approaching a limiting value which called the maximum gravitational mass.  Furthermore, when the $\lambda$ value decreases, the masses and the corresponding radii of the NS increase.  We see that for $\lambda =-3$, maximum mass reaches to $2.509M_{\odot}$ (see Table \ref{results1} for more details). As illustrated in Fig. \ref{6}, from $\lambda =-0.001$ to $\lambda =-2$, Our curves show good agreement with the observational data for {pulsars PSR J0952-0607 and PSR J2215+5135.} This indicates that even in the case of isotropic NSs, the aforementioned EoS  in the framework of $f(\mathcal{R},T)$, theoretical results could potentially account for unexplained masses within the mass gap.

\begin{figure}[h!]
    \centering
    \includegraphics[width=8.5cm]{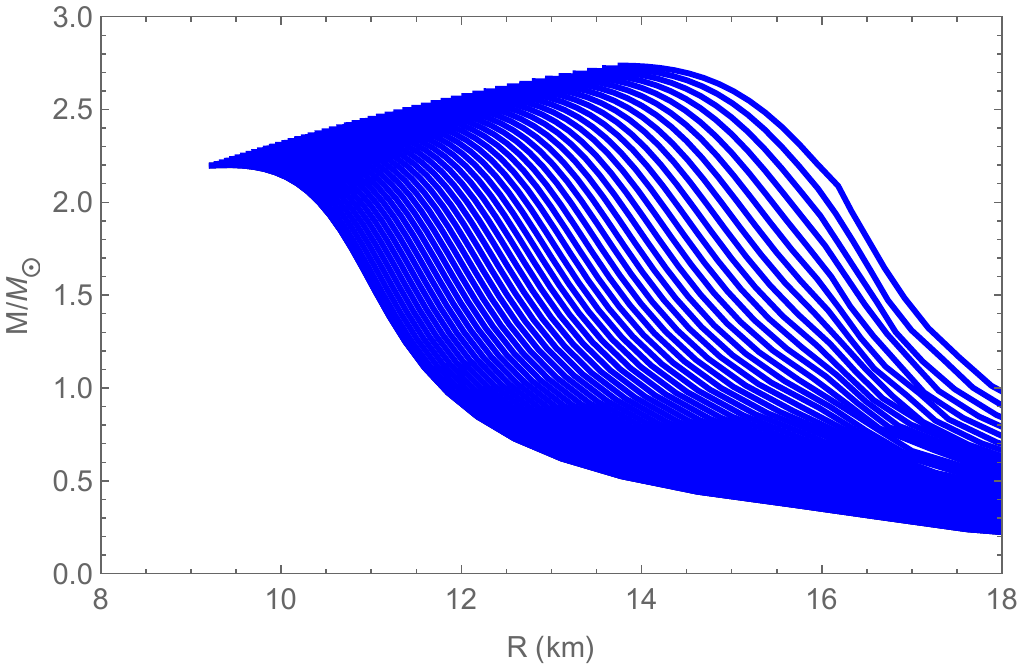}
    \caption{Monotonic behavior of the linear madel of $f(\mathcal{R},T)$ gravity with 50 coupling parameter ($\lambda$) data points. } \label{5}
\end{figure}

 Furthermore, One of the most important features of studying $f(\mathcal{R},T)$ models is to see that there are no irregular jumps of the coupling parameters ($\lambda$) in specific regime, like the one which we investigate in this manuscript (see Fig. \ref{5}). As it is seen, clearly there is no irregular jumps in the $\lambda$ parameters. The objective of Fig. \ref{5}   is to ensure that you feel at ease with these $\lambda$ parameters, as they have been meticulously selected.   To demonstrate that the linear gravity model exhibits monotonic behavior, we have plotted the mass-radius relations for values of the coupling parameter from $\lambda=-10^{-4}$ to $\lambda=-5$ corresponding to 50 coupling parameter data points. When  $\lambda$ parameter tends to $-\frac{8\pi}{3}$; that is, if our gravity model becomes $f(\mathcal{R},T)=\mathcal{R}-\frac{16\pi}{3} T$, the TOV equations become unsolvable. In fact, the Schwarzschild radius vanishes, and we can no longer study compact stars with this model. In this limit every compact object would become a black hole.
We set the parameters such that the model simultaneously traverses the constraints from mass and radius of the component of GW170817,  and masses of the components of GW190814 and GW200210-092254, as well as satisfies the constraints from  masses and radii of PSR J0740+6620 and PSR J0030+0451 and also masses of PSR J2215+5135 and PSR J0952-0607. Consequently, the model successfully fulfills all relevant observational constraints concurrently. Our calculations indicate that to comprehensively encompass all the aforementioned constraints, the free parameters must be approximately $\beta= 0.2 - 1.0$ and  $\lambda  $ from $-0.1$  to  $-3.0$. This is why in this study we select $\lambda$ and  $\beta$ values only in the aforementioned ranges, despite theoretically permissible values of $\lambda$ up to $-\frac{8\pi}{3}$.

\subsection{Modified Anisotropic TOV Equations in $f(\mathcal{R},T)$
Gravity} \label{SubSec.IIIB}

We model the matter distribution within the NS as an anisotropic fluid, characterized by the energy density $\epsilon$, radial pressure $P_{r}$, and tangential pressure $P_{t}$. In this case, the corresponding energy-momentum tensor takes the form \cite{Pretel:2022qng-69},
\begin{eqnarray}
T_{\mu \nu }=(\epsilon c^2 +P_{t}) u_{\mu}u_{\nu} + P_{t} g_{\mu \nu} - \Delta V_{\mu} V_{\nu},    \label{31}
\end{eqnarray}
where $u_{\mu}$ defined  benight Eq. (\ref{28}), $V_{\mu}$ is a unit radial four-vector satisfying $V_{\mu}V^{\mu}=1$ and $\Delta $ quantifies the pressure anisotropy.  In the relevant equations in the context of anisotropic TOV solutions within $f(\mathcal{R},T)$ gravity, we require a well-defined functional form for anisotropy. This formalism must account for both the local anisotropy of the fluid and the influence of strong magnetic field. While microscopic theories provide insights into anisotropy, there remains no universally accepted method to explicitly describe the combined effects of fluid anisotropy and magnetic field in compact objects \cite{Bordbar:2024yai-71}. 
Various models exist for the anisotropy function 
$\Delta$, such as Horvat model \cite{horvat11-1045} , Herrera-Barreto model \cite{herrera13-084022} and so on \cite{thirukkanesh08-235001, biswas19-104002, gokhroo94-75}; however, it is not our objective to investigate all of them. Among these anisotropy prescriptions, we select only the Bowers-Liang model \cite{BowerLiang-48}, as it is one of the most well-known in the literature \cite{mak09-aniso, raghuraman21-bowers, deb21-bowers} and references therein. Our aim is to examine the effect of this particular type (i.e. Bowers-Liang model) of anisotropy on the structure of NSs, although a more comprehensive analysis would require exploring various parameterizations of $\Delta$.

\begin{figure*}
    \includegraphics[width=8.9cm]{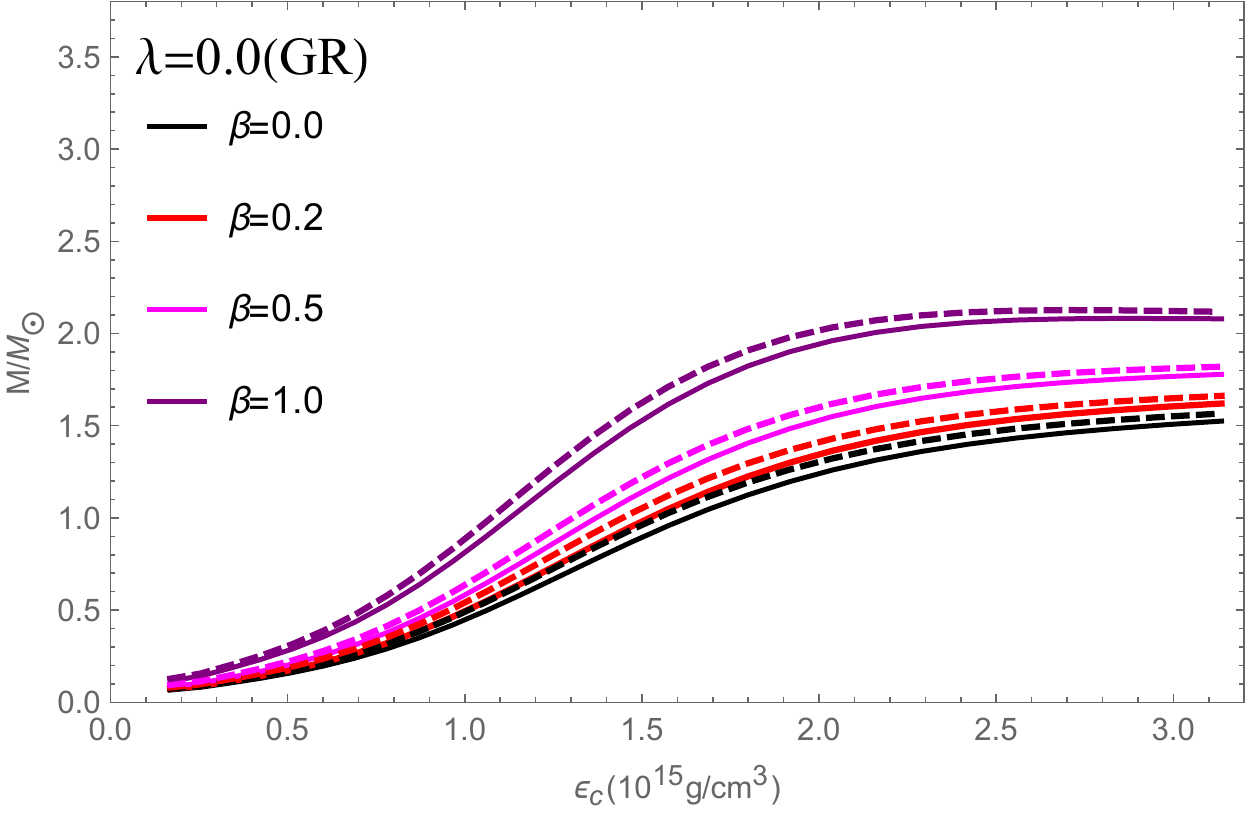}
    \includegraphics[width=8.9cm]{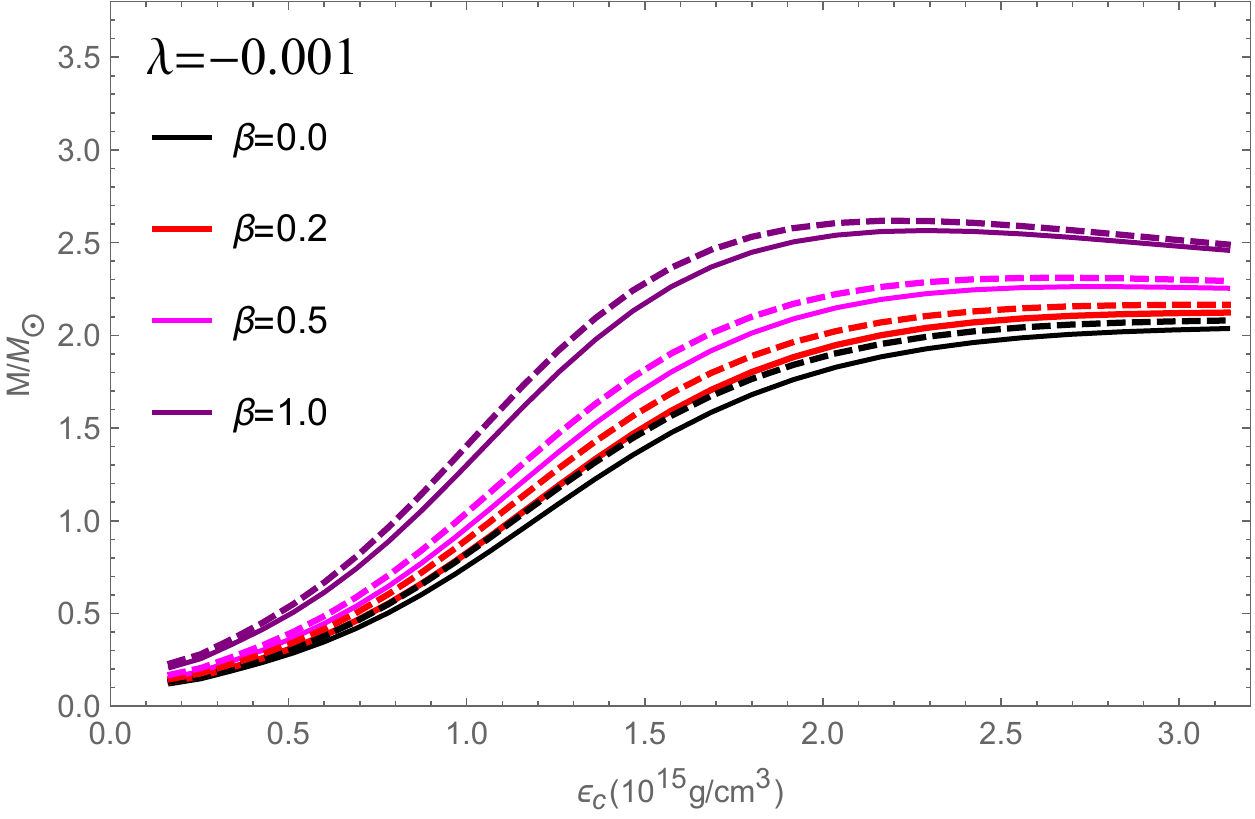}\\
    \includegraphics[width=8.9cm]{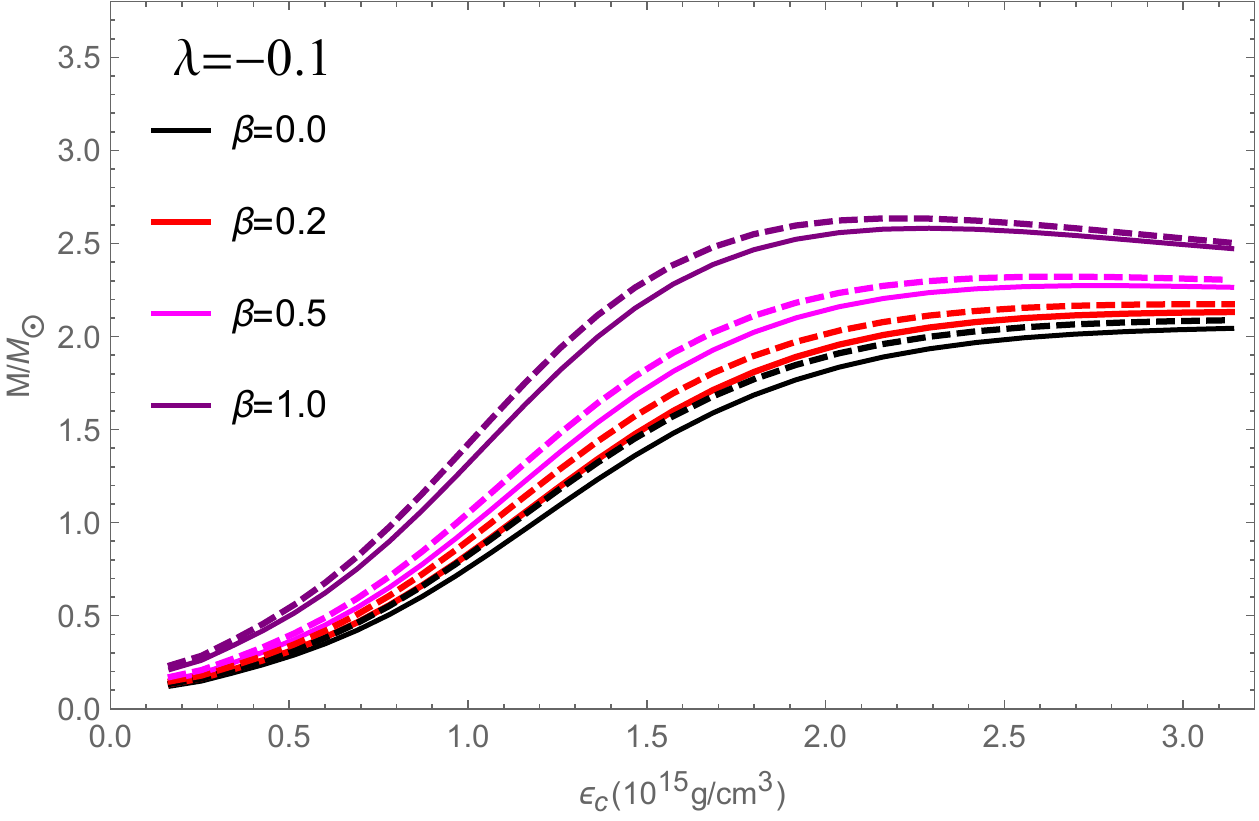}
    \includegraphics[width=8.9cm]{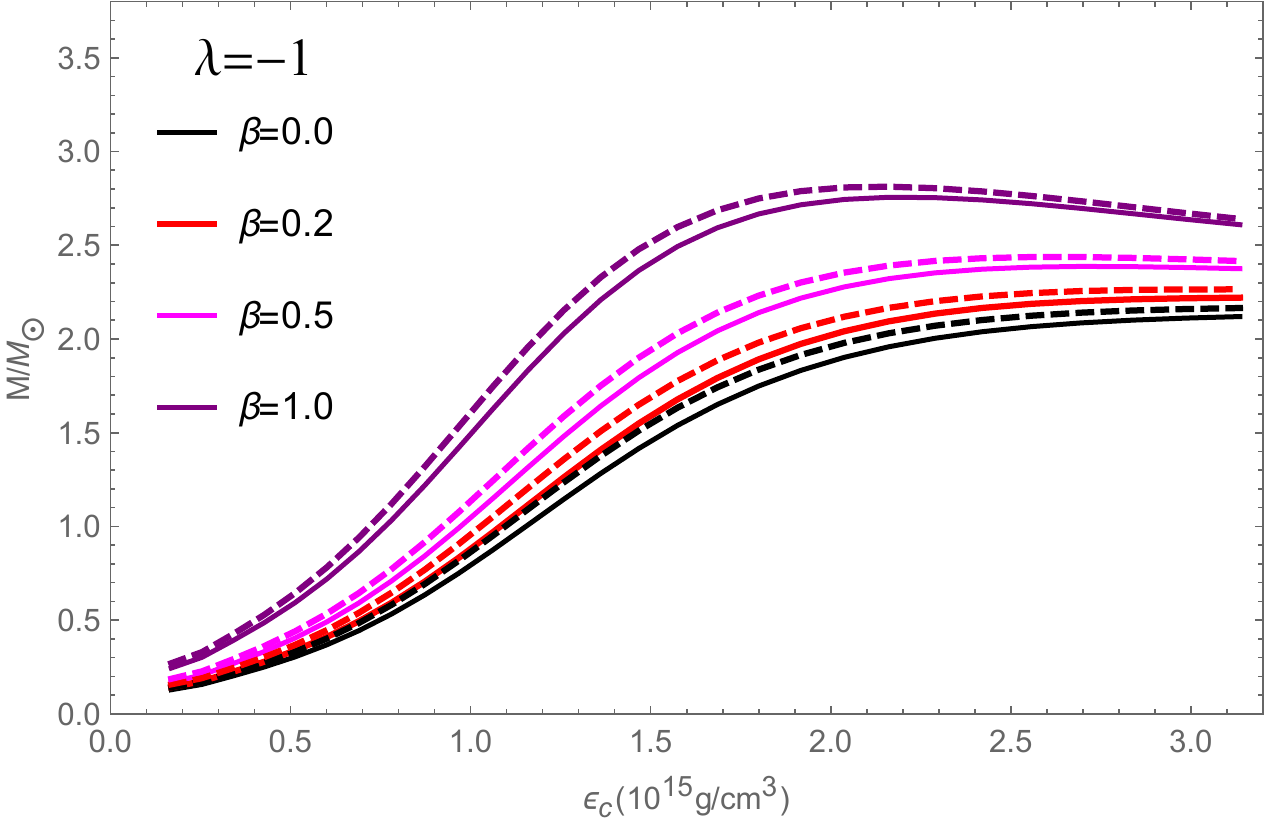}\\
    \includegraphics[width=8.9cm]{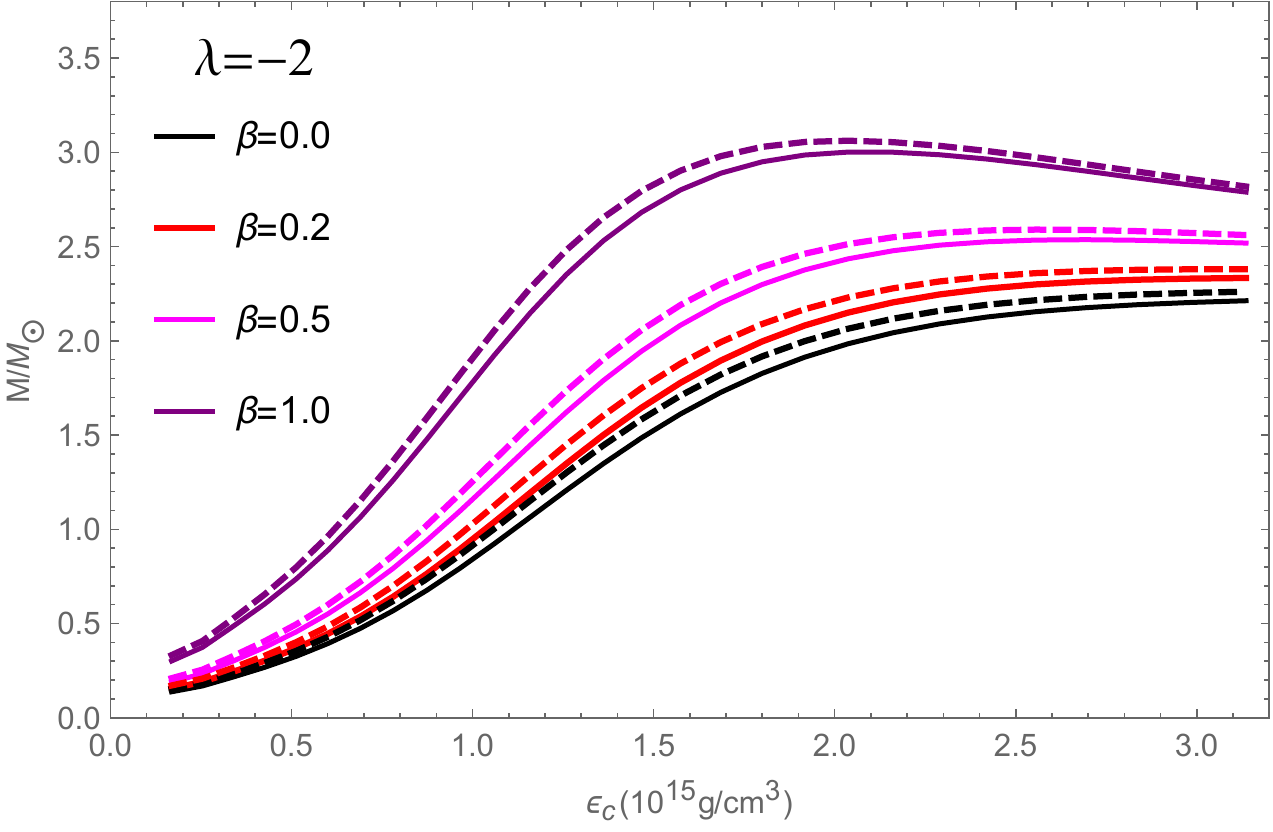}
    \includegraphics[width=8.9cm]{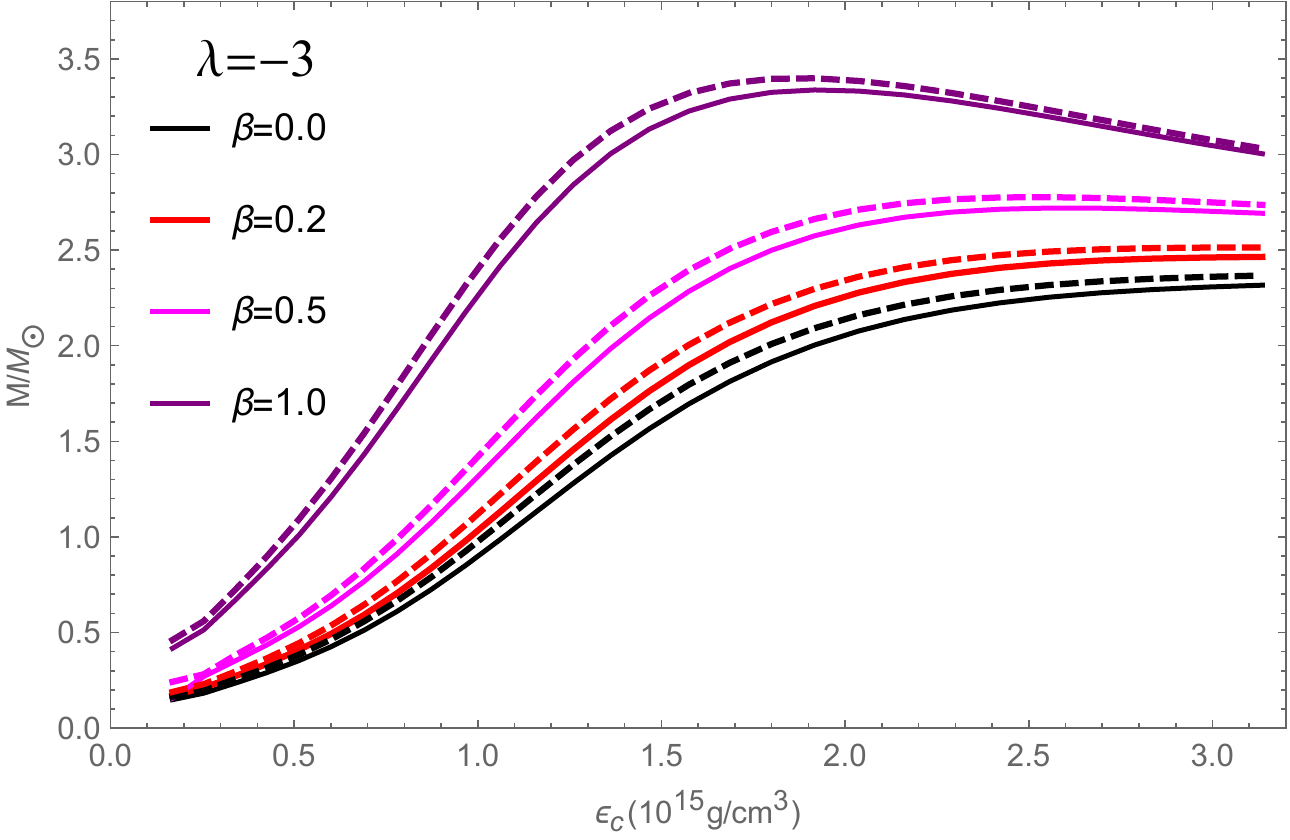}\\
    \caption{The mass-central energy density panels for two  magnetic fields $B_{surf}=5.0 \times 10^{16} G$ (dashed) and $B_{surf}=1.0 \times 10^{17} G$ (solid lines) for $B_{0}=2.0\times 10^{18} G $ and for different coupling parameters ($\lambda$). As demonstrated, with decreasing values of $\lambda$ leading to an increase in  the maximum masses  of the NSs.  }
    \label{121}
\end{figure*}

\begin{figure*}
    \includegraphics[width=8.9cm]{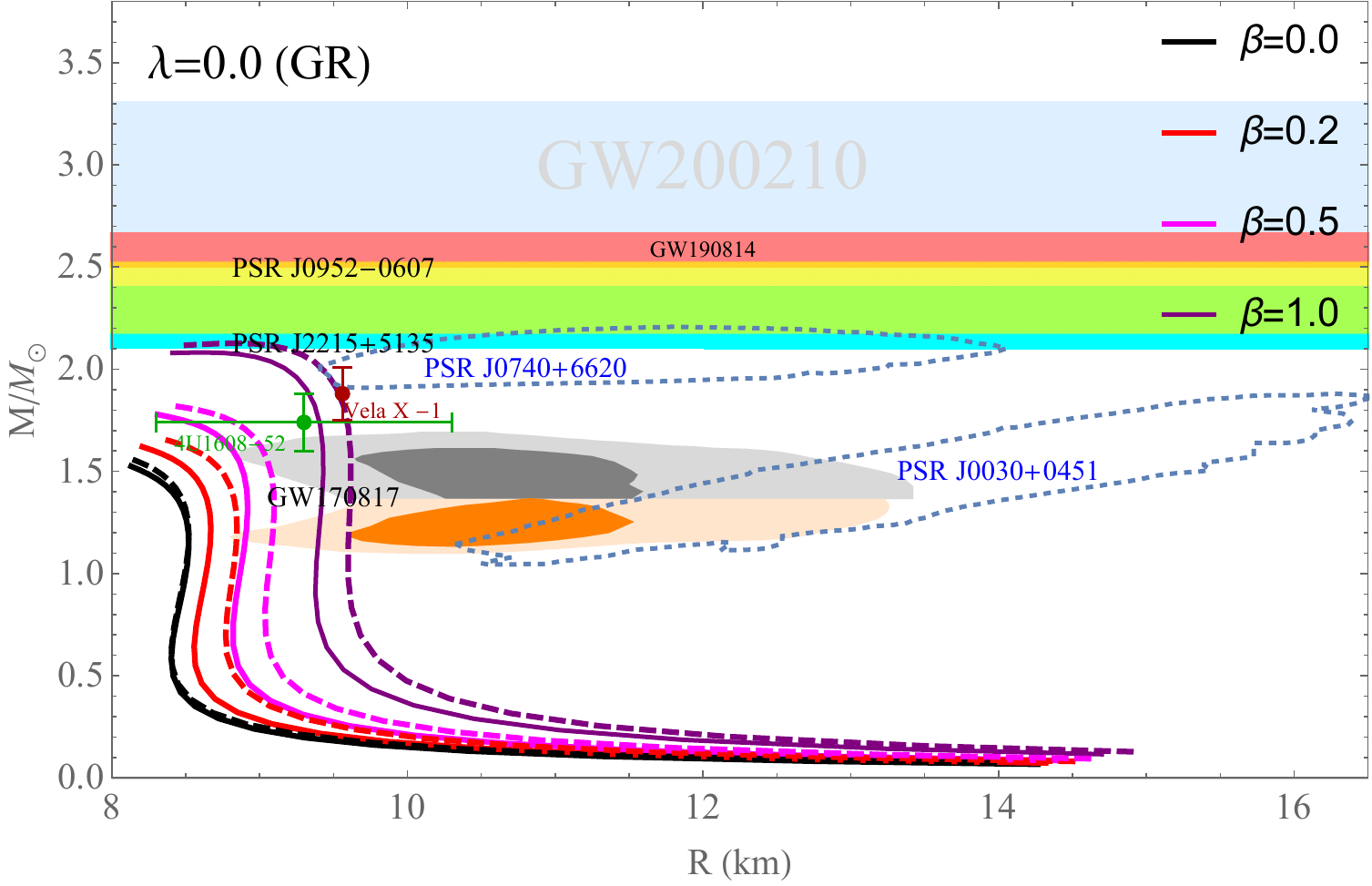}
    \includegraphics[width=8.9cm]{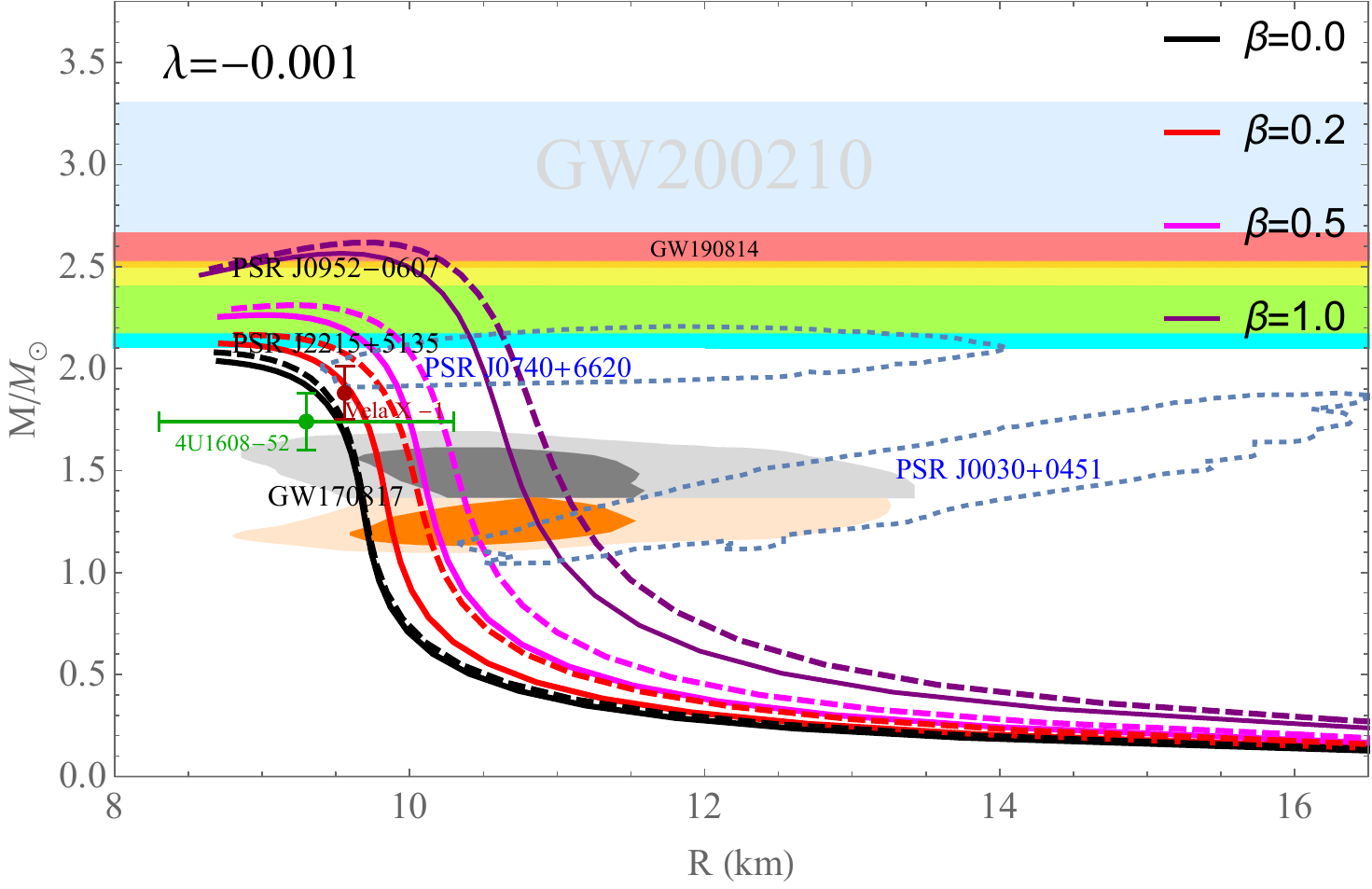}\\
    \includegraphics[width=8.9cm]{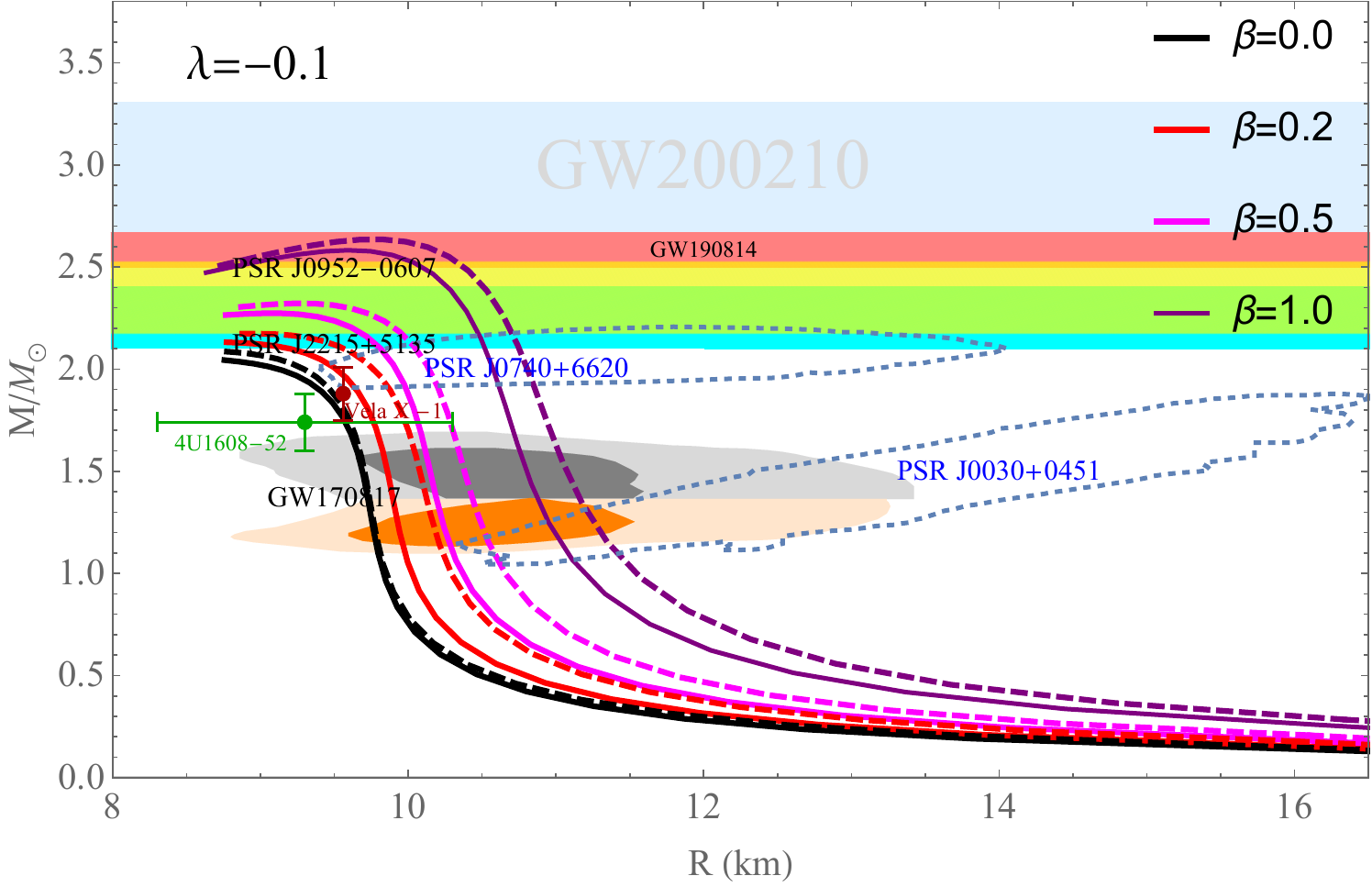}
    \includegraphics[width=8.9cm]{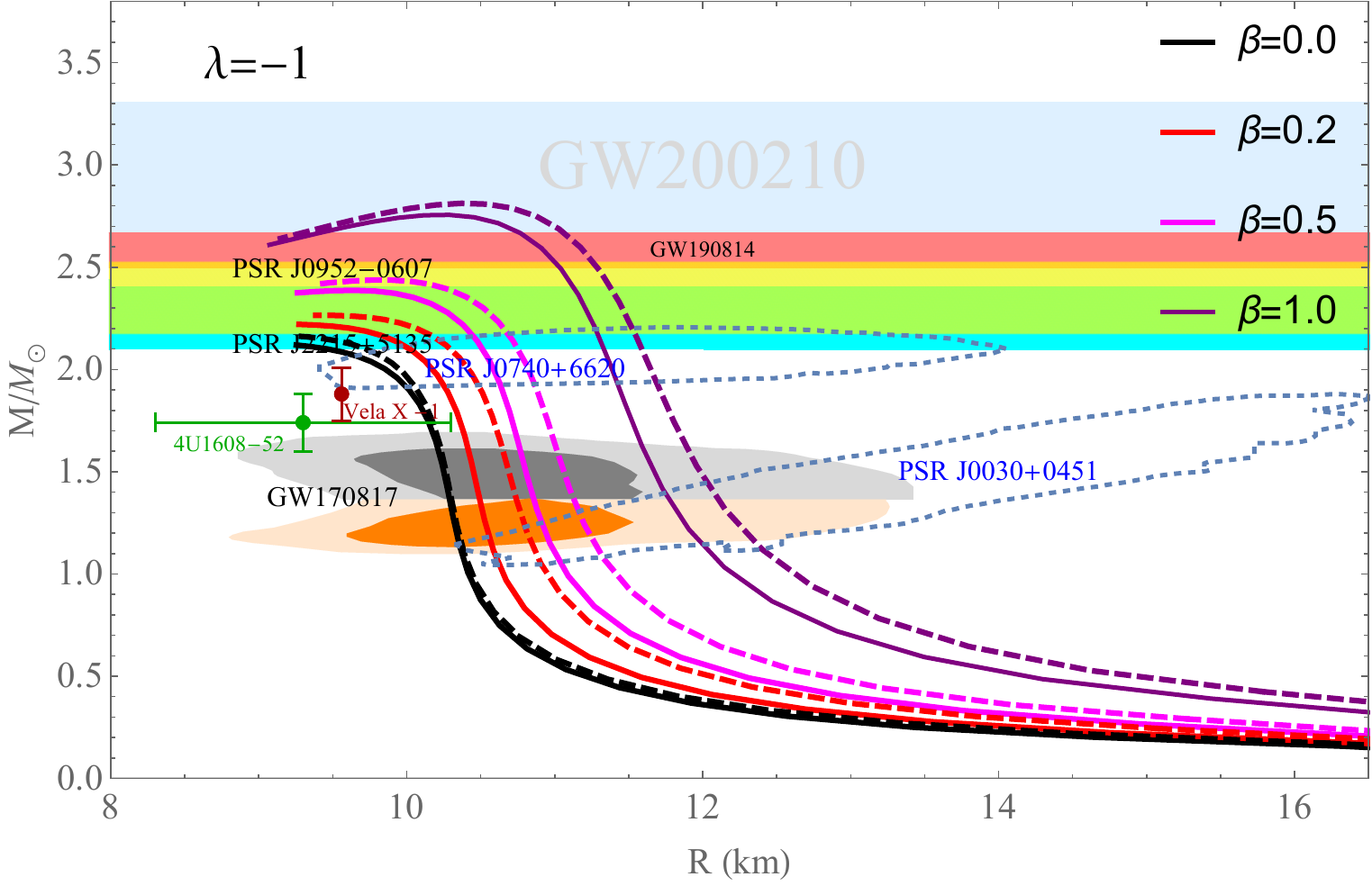}\\
    \includegraphics[width=8.9cm]{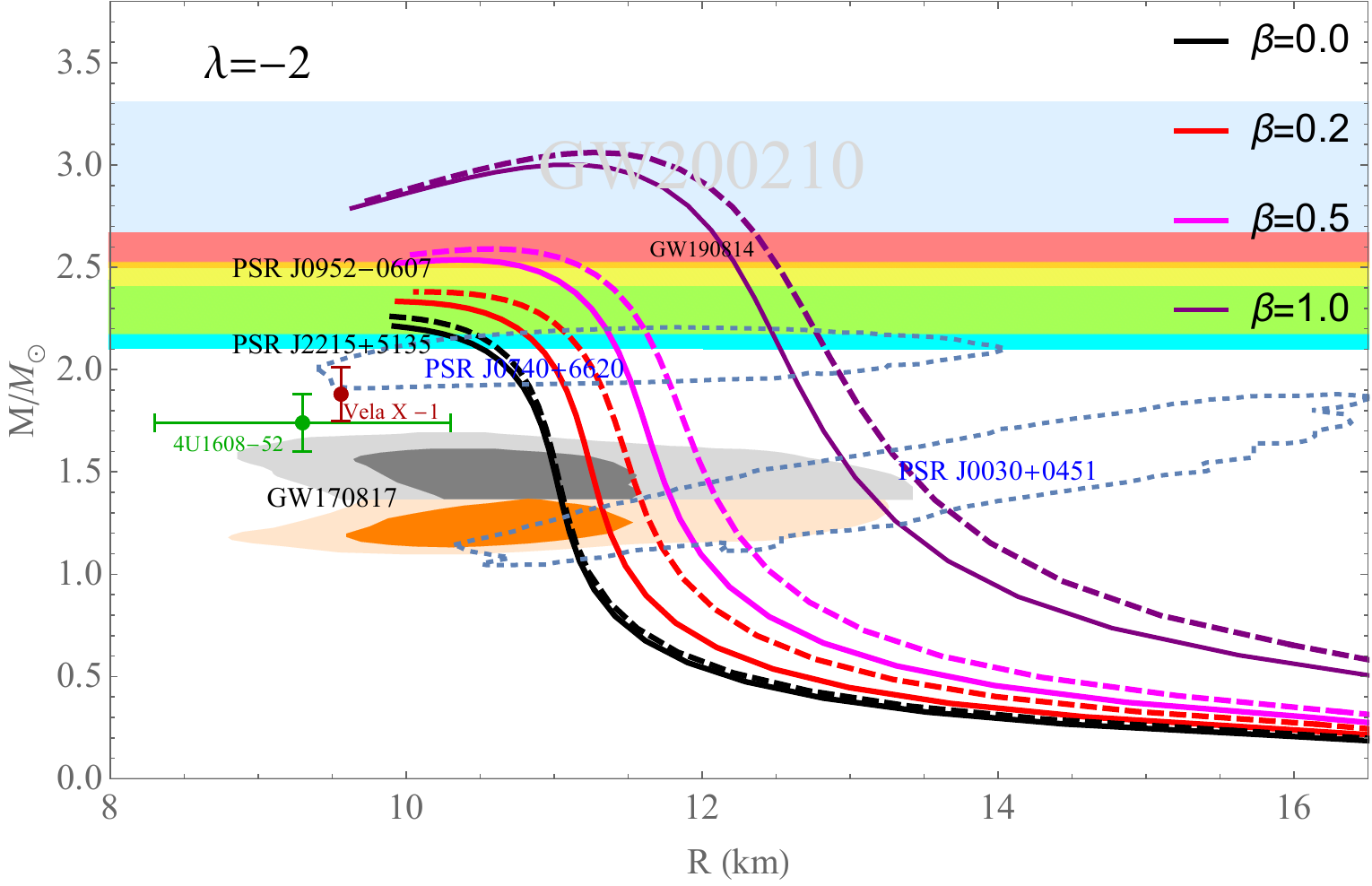}
    \includegraphics[width=8.9cm]{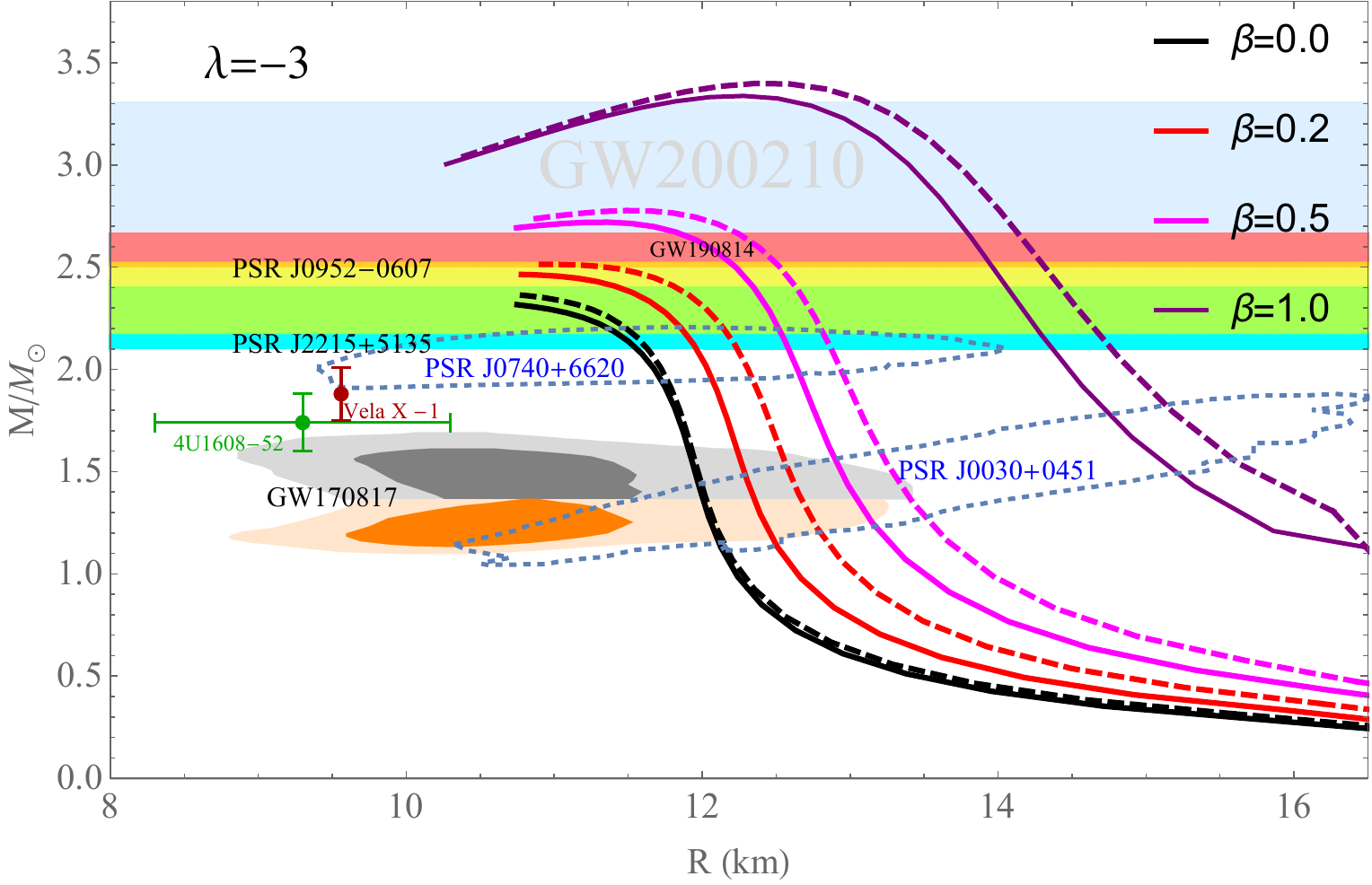}\\
    \caption{The mass-radius relations for two  magnetic fields $B_{surf}=5.0 \times 10^{16} G$ (dashed) and $B_{surf}=1.0 \times 10^{17} G$ (solid lines) for $B_{0}=2.0\times 10^{18} G $ and for different coupling parameters ($\lambda$). As demonstrated, with decreasing values of $\lambda$ leading to an increase in both the maximum masses and corresponding radii of the NSs. The mass of the {pulsar PSR J0952-0607 (yellow), the mass of the pulsar PSR J2215+5135} (cyan), and their overlapping region (green) are shown. Also the central hatched area corresponds to the mass of the component of GW170817  and the light blue region corresponds to the mass of GW200210-092254 . }
    \label{12}
\end{figure*}

Thus, in line with the approach of Bowers and Liang \cite{BowerLiang-48}, the anisotropy parameter  is defined as follows,
\begin{equation}
\Delta=\frac{\beta G}{c^{4}}(\epsilon c^2+ P_{r})({\epsilon c^2+3P_{r}})e^{\Psi}r^{2},
\label{32}
\end{equation}
where $\beta$ is the Bowers-Liang's parameter and $e^{\Psi}$ is defined in Eq. (\ref{25}).
\begin{table*}
\caption{Maximum masses and corresponding radii of anisotropic magnetized NS for different coupling parameters ($\lambda$) in $f(R,T)$ gravity.}
\label{IIIX}
\begin{tabular*}{\textwidth}{@{\extracolsep{\fill}}lrrr|rl@{}}
\hline
\hline
$B_{surf} $ &$ $ & \multicolumn{1}{c}{ $ 5.0 \times 10^{16} [G]$} & \multicolumn{1}{c}{$$} & \multicolumn{1}{c}{$1.0 \times 10^{17} [G]$} & \multicolumn{1}{c}{ $$} \\
\hline
$\lambda $ &$\beta $ & \multicolumn{1}{c}{ ${M_{max}}\ [M_{\odot}]$} & \multicolumn{1}{c}{$R\ [km]$} & \multicolumn{1}{c}{${M_{max}}\ [M_{\odot}]$} & \multicolumn{1}{c}{ $R\ [km]$} \\
\hline
           $$&$0.0$ & $1.524$ & $8.130$ & $1.431$ & $7.805$ \\
           $$&$0.1 $ & $1.571$ & $8.168$ & $1.477$ & $7.847$ \\
           $$&$0.2$ & $1.620$ & $8.206$ &  $1.526$ & $7.888$  \\
$0.0(GR)$&$0.3$ & $1.670$ & $8.242$ & $1.576$ & $7.928$ \\
           $$&$0.5$ & $1.778$ & $8.309$ & $1.684$ & $8.004$ \\
           $$&$0.8$ & $1.954$ & $8.385$ & $1.864$ & $8.101$ \\
           $$&$1.0$ & $2.081$ & $8.639$ & $1.994$ & $8.145$ \\

\hline
          $$&$0.0$ & $2.036$ & $8.706$ & $1.934$ & $8.401$ \\
          $$&$0.1 $ & $2.079$ & $8.714$ & $1.977$ & $8.415$ \\
          $$&$0.2$ & $2.122$ & $8.720$ & $2.022$ & $8.427$ \\
$-0.001$&$0.3$ & $2.166$ & $8.723$ & $2.067$ & $8.435$\\
          $$&$0.5$ & $2.262$ & $8.963$ & $2.159$ & $8.442$ \\
          $$&$0.8$ & $2.432$ & $9.378$ & $2.307$ & $8.832$ \\
          $$&$1.0$ & $2.565$ & $9.537$ & $2.428$ & $8.982$  \\

\hline
   $$&$0.0$ & $2.212$ & $9.919$ & $2.098$ & $9.562$ \\
   $$&$0.1 $ & $2.272$ & $9.933$ & $2.159$ & $9.583$ \\
   $$&$0.2$ & $2.333$ & $9.942$ & $2.221$ & $9.599$ \\
$-2$&$0.3$ & $2.395$ & $10.070$ & $2.285$ & $9.610$ \\
   $$&$0.5$ & $2.536$ & $10.350$ & $2.414$ & $9.610$ \\
   $$&$0.8$ & $2.794$ & $10.828$ & $2.646$ & $10.204$ \\
   $$&$1.0$ & $3.001$ & $11.188$ & $2.839$ & $10.554$  \\

\hline
   $$&$0.0$ & $2.316$ & $10.748$ & $2.194$ & $10.356$ \\
   $$&$0.1 $ & $2.389$ & $10.767$ & $2.269$ & $10.382$ \\
   $$&$0.2$ & $2.464$ & $10.778$ &$2.345$ & $10.402$ \\
$-3$&$0.3$ & $2.541$ & $10.914$ & $2.432$ & $10.414$ \\
   $$&$0.5$ & $2.719$ & $11.366$ & $2.584$ & $10.553$ \\
   $$&$0.8$ & $3.055$ & $11.896$ & $2.891$ & $11.219$ \\
   $$&$1.0$ & $3.337$ & $12.282$ & $3.156$ & $11.591$  \\
\hline
\hline
\end{tabular*}
\end{table*}
This anisotropy parameter ensures compatibility with the anisotropic modified  TOV equations while preserving the physical constraints imposed by NS structure. By setting $\beta=0$, we return to the isotropic state (see  Ref. \cite{BowerLiang-48} and Subsec. \ref{SubSec.IIIA} for further discussion). Thus, we adopt the EoS to solve the governing hydrostatic equilibrium equations and derive the NS's properties.  This formulation explicitly separates the isotropic $P_{t}$
 and anisotropic $\Delta$ contributions to the energy-momentum tensor, enabling a systematic analysis of anisotropic effects in compact objects. In addition, we consider the line element in Eq. (\ref{21}), and the trace of the energy-momentum tensor (Eq. (\ref{31})) takes the form  $T=-\epsilon c^2+3P_{r}+2\Delta$. Within the framework of anisotropic fluids in $f(R,T)$ gravity, the most adopted choice in the literature for the matter Lagrangian density is given by $\mathcal{L}_m = \mathcal{P}$, where $\mathcal{P} \equiv (P_r+ 2P_t)/3$. Under this consideration, $\Theta_{\mu\nu} = -2T_{\mu\nu} + \mathcal{P}g_{\mu\nu}$ the system of dynamical equations will be as follow \cite{Pretel:2022qng-69},
\begin{align}
    G_{\mu\nu} &= \frac{8\pi G}{c^4} T_{\mu\nu} + \lambda Tg_{\mu\nu} + 2 \lambda(T_{\mu\nu} - \mathcal{P}g_{\mu\nu}) ,   \label{33}   \\
    \mathcal{R} &= \frac{-8\pi G}{c^4} T - 2 \lambda(3T- 4\mathcal{P}) ,   \label{34}    \\
    \nabla^\mu T_{\mu\nu} &= \frac{2 \lambda}{8\pi  + 2 \lambda} \partial_\nu\left( \mathcal{P} - \frac{1}{2}T \right) .   \label{35}
\end{align}
For the metric in Eq. (\ref{21}) and energy-momentum tensor in Eq. (\ref{31}), the non-zero components of the field equations  are explicitly given by,
 \begin{align}
      & \frac{1}{r^2}\frac{d}{dr}(re^{-\Psi}) - \frac{1}{r^2} = \frac{-8\pi G}{c^2}\epsilon +  \lambda\left[ -3\epsilon c^2+P_{r} + \frac{2}{3}\Delta \right] ,  \label{36}  \\
      & e^{-\Psi}\left( \frac{1}{r}\Phi' + \frac{1}{r^2} \right)- \frac{1}{r^2} = \frac{8\pi G}{c^4} P_{r} +  \lambda\left[ -\epsilon c^2+ 3P_{r}+ \frac{2}{3}\Delta \right].  \label{37}
      %
  \end{align}
With some calculations, we can obtain the TOV equations in an anisotropic scenario of the $f(\mathcal{R},T)$ modified gravity as follows,
\begin{align}
    \frac{dm}{dr} =&\ 4\pi r^2\epsilon c^2+ \frac{\lambda r^2}{2}\left( 3\epsilon c^2-P_{r}- \frac{2}{3}\Delta \right) ,  \label{39}  \\
    \frac{dP_r}{dr} =& -\frac{\epsilon c^2+ P_{r}}{1+\zeta} \left[ \frac{Gm}{c^2r^2}+ 4\pi r P_{r} + \frac{\lambda r}{2}\left( 3P_{r} - \epsilon c^2+ \frac{2}{3}\Delta \right) \right]  \nonumber  \\
    &\times \left( 1- \frac{2Gm}{rc^2} \right)^{-1} + \frac{\zeta c^2}{1+\zeta}\frac{d\epsilon}{dr}
    + \frac{2}{1+\zeta}\left[ \frac{\Delta}{r} - \frac{\zeta}{3}\frac{d\Delta}{dr} \right] ,  \label{40}
\end{align}
Eqs. (\ref{39}) and  (\ref{40}) are the set of equations that are called "modified anisotropic TOV equations in $f(\mathcal{R},T)$
gravity". As expected, the modified TOV equations in the anisotropic case are
retrieved when $P_{r}=P_{t}$, and  when the minimal coupling parameter vanishes $\lambda =0$, we can
recover the standard TOV equations for isotropic NSs
in GR. By solving Eqs. (\ref{39}) and  (\ref{40}) numerically, the structure parameters of an anisotropic compact object can be computed.
Our results for the anisotropic case of neutron star are presented in Figs. \ref{121} and  \ref{12}   for $B_{surf}=5.0\times 10^{16}G$ and  $B_{surf}=1.0\times 10^{17}G$.
 The results demonstrate that the mass-radius relations exhibits a strong dependence on the coupling parameter $\lambda$. Here we see that  decreasing the value of $\lambda$ and increasing $\beta$ leading to an increase in both the maximum mass and corresponding radius of the NS. In fact, as the EoS  becomes softer in stronger surface magnetic fields (see Fig. \ref{2}), it predicts lower masses with increasing anisotropy parameter.  It should be noted that as the $\beta$ parameter increases, i.e., with increasing stellar anisotropy, the mass increases.
Increasing the surface magnetic field raises the anisotropy
parameter, but softens the EoS, leading to a net decrease in mass.
From Fig. \ref{12}, for $\lambda=0$, we see that for high surface
magnetic field $ 1.0 \times 10^{17} G $ the results do not show
agreement with observational data, whereas the results of the
$f(\mathcal{R},T)$ theory are consistent with those of
observations, specially for $\lambda$ values in this range $-0.1$
to $-3$.

\subsection{Effects of Parameter $\beta$ on magnetized  Neutron Star's Structure} \label{Sec.IV}
\begin{figure}[h!]
    \centering
    \includegraphics[width=8.5cm]{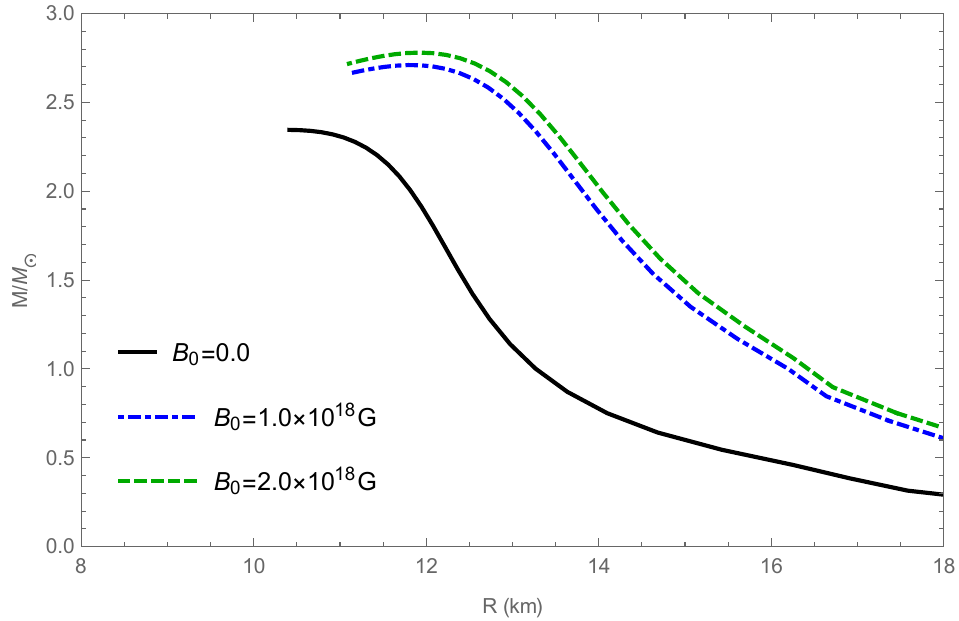}
    \caption{The effect of the central magnetic field on the mass-radius relation at $\lambda=-2.0$ and $B_{surf}=5\times10^{16}G$.}    \label{9}
\end{figure}

For selected values of the coupling parameter ($\lambda$) at
$B_{0}=2.0\times 10^{18} G $ and $B_{surf}=5\times10^{16}G$, we
investigate the influence of the parameter $\beta$  on the
mass-radius relations within the framework of $f(\mathcal{R},T)$
gravity. Furthermore, we analyze the impact of the anisotropy
factor on  structural properties of NSs, including their maximum
mass, radius, etc. To validate our theoretical predictions, we
compare our computational results with established observational
data
\cite{Romani:2022jhd-72,romani12-73,Miller21-74,Zhao15-75,Kretschmar:2019hyk-76,Bhattacherjee24-77,Lenzi:2013dya-78}
(see Table \ref{XI}), thereby incorporating observational
constraints into the modeling of anisotropic NS structure in
$f(\mathcal{R},T)$ gravity. Our analysis reveals that achieving
consistency between theoretical calculations and observational
measurements requires restricting the anisotropy factor to a
specific range (see Table \ref{II} for more detail). For a fixed
coupling parameter ($\lambda$), we arbitrarily select $\beta$
values ranging from zero to one, and observe that at any
$\lambda$, the masses and corresponding radii increase with
increasing $\beta$. As evident in Fig. \ref{12}, we have found
that for $\beta$  beyond unity, the resulting masses correspond to
none of the observed compact objects. Therefore, such large masses
compromise stellar stability, as they overwhelm the degeneracy
pressure of neutron matter, leading to gravitational collapse into
a black hole. In addition to the stiffness of the EoS, the
anisotropy parameter itself plays a significant role.
Specifically, lower $\beta$ values yield lower masses, whereas
higher $\beta$ values (even with a soft EoS) yield larger masses
(refer to Table \ref{IIIX}). Furthermore, a comparison of the
various panels in Fig.  \ref{12} and Tables  \ref{II} and \ref{VI}
reveals that for fixed $\beta$, a decreasing in the coupling
parameter ($\lambda$) leads to an increase in the maximum masses
and  corresponding radii. For instance, for $\beta=1.0$ and
magnetic field $1.0 \times 10^{17}G$, we obtained a mass and
radius $2.755\ M_{\odot}$ and $10.288\ km$  for $\lambda=-1.0$,
and a mass and radius $2.839\ M_{\odot}$ and $10.554\ km$  for
$\lambda=-2.0$, respectively.
\begin{table*}
\caption{Properties of isotropic non-magnetized NSs for different coupling parameters in $f(\mathcal{R},T)$ gravity.}
\label{I}
\begin{tabular*}{\textwidth}{@{\extracolsep{\fill}}lrrrrl@{}}
\hline
\hline
$\lambda $ &  \multicolumn{1}{c}{$R_{Sch}\ [km]$} & \multicolumn{1}{c}{ $C$} & \multicolumn{1}{c}{$z$} & \multicolumn{1}{c}{$K[10^{-8}$ $m^{-2}]$} &  \multicolumn{1}{c}{$d(^{0}/_{0})$} \\
\hline
$0.0(GR)$ &  $4.966$ & $0.284$ & $0.271$ & $2.587$ & $00.00$\\
$-10^{-7} $ &  $5.167$ & $0.293$ & $0.279$ & $2.616$& $+01.12$\\
$-0.001$ &  $6.486$ & $0.338$ & $0.319$ & $2.557$ & $-01.15$\\
$-0.5$ & $6.231$ & $0.323$ & $0.299$ & $1.312$ & $-49.28$\\
$-1.0$ &  $5.957$ & $0.299$ & $0.272$ & $2.587$ & $00.00$\\
$-2.0$ & $5.381$ & $0.251$ & $0.212$ & $2.009$& $-22.34$ \\
$-3.0$ & $4.775$ & $0.204$ & $0.147$ & $3.286$& $+27.01$ \\
\hline
\hline
\end{tabular*}
\end{table*}

\begin{table*}
\caption{Properties of anisotropic magnetized neutron stars for  $B_{surf}=5.0\times 10^{16} G$ and $B_{0}=2.0\times 10^{18} G $.}
\label{II}
\begin{tabular*}{\textwidth}{@{\extracolsep{\fill}}lrrrrrl@{}}
\hline
\hline
$\lambda $ &$\beta $ & \multicolumn{1}{c}{$R_{Sch}\ [km]$} & \multicolumn{1}{c}{ $C$} & \multicolumn{1}{c}{$z$} & \multicolumn{1}{c}{$K[10^{-8}$ $m^{-2}]$} &  \multicolumn{1}{c}{$d(^{0}/_{0})$} \\
\hline
$$&$0.0$ &  $4.500$ & $0.276$ & $0.264$ & $2.901$ & $00.00$\\
$$&$0.1 $ &  $4.639$ & $0.283$ & $0.270$ & $2.949$ & $00.00$\\
$$&$0.2$ &  $4.783$ & $0.291$ & $0.277$ & $2.99$ & $00.00$\\
$0.0(GR)$&$0.3$&   $4.931$ & $0.299$ & $0.284$ & $3.051$  & $00.00$\\
$$&$0.5$ &   $5.250$ & $0.315$ & $0.299$ & $3.170$  & $00.00$\\
$$&$0.8$ &   $5.770$ & $0.344$ & $0.324$ & $3.390$  & $00.00$\\
$$&$1.0$ &   $6.146$ & $0.355$ & $0.334$ & $3.301$ & $00.00$ \\

\hline
$$&$0.0$ &  $6.011$ & $0.345$ & $0.325$ & $2.359$ & $-22.97$\\
$$&$0.1 $ & $6.138$ & $0.352$ & $0.331$ & $2.425$ & $-21.60$\\
$$&$0.2$ &  $6.265$ & $0.359$ & $0.337$ & $2.496$ & $-20.15$\\
$-0.001$&$0.3$ &  $6.395$ & $0.366$ & $0.344$ & $2.570$  & $-18.71$\\
$$&$0.5$ &  $6.678$ & $0.372$ & $0.349$ & $2.522$  & $-25.69$\\
$$&$0.8$ &   $7.180$ & $0.382$ & $0.358$ & $2.420$  & $-40.08$\\
$$&$1.0$ &   $7.573$ & $0.397$ & $0.371$ & $2.451$ & $-34.67$ \\

\hline
$$&$0.0$ &   $4.972$ & $0.250$ & $0.211$ & $3.040$ & $+04.57$\\
$$&$0.1 $ &   $5.107$ & $0.257$ & $0.216$ & $3.103$ & $+04.96$\\
$$&$0.2$ &   $5.244$ & $0.263$ & $0.221$ & $3.171$ & $+05.42$\\
$-2$&$0.3$ &   $5.384$ & $0.267$ & $0.224$ & $3.125$  & $+02.36$\\
$$&$0.5$ &   $5.700$ & $0.275$ & $0.230$ & $3.039$  & $-04.31$\\
$$&$0.8$ &   $6.280$ & $0.290$ & $0.240$ & $2.928$  & $-15.77$\\
$$&$1.0$ &   $6.746$ & $0.301$ & $0.249$ & $2.864$ & $-15.25$ \\

\hline
$$&$0.0$ &   $4.390$ & $0.202$ & $0.146$ & $4.469$ & $+35.08$\\
$$&$0.1 $ &   $4.528$ & $0.210$ & $0.150$ & $4.586$ & $+35.69$\\
$$&$0.2$ &   $4.670$ & $0.216$ & $0.154$ & $4.716$ & $+36.40$\\
$-3$&$0.3$ &   $4.816$ & $0.220$ & $0.157$ & $4.684$  & $+34.86$\\
$$&$0.5$ &   $5.153$ & $0.226$ & $0.160$ & $4.444$  & $+28.66$\\
$$&$0.8$ &   $5.790$ & $0.243$ & $0.170$ & $4.382$  & $+22.63$\\
$$&$1.0$ &    $6.325$ & $0.257$ & $0.178$ & $4.378$ & $+24.66$ \\
\hline
\hline
\end{tabular*}
\end{table*}

\begin{table*}
\caption{Properties of anisotropic magnetized neutron stars for  $B_{surf}=1.0\times 10^{17} G$ and $B_{0}=2.0\times 10^{18} G $ .}
\label{VI}
\begin{tabular*}{\textwidth}{@{\extracolsep{\fill}}lrrrrrl@{}}
\hline
\hline
$\lambda $ &$\beta $ &  \multicolumn{1}{c}{$R_{Sch}\ [km]$} & \multicolumn{1}{c}{ $C$} & \multicolumn{1}{c}{$z$} & \multicolumn{1}{c}{$K[10^{-8}$ $m^{-2}]$} &  \multicolumn{1}{c}{$d(^{0}/_{0})$} \\
\hline

$$&$0.0$ &  $4.225$ & $0.270$ & $0.258$ & $3.079$ & $00.00$\\
$$&$0.1 $ &  $4.361$ & $0.277$ & $0.265$ & $3.127$ & $00.00$\\
$$&$0.2$ &   $4.506$ & $0.285$ & $0.272$ & $3.180$ & $00.00$\\
$0.0(GR)$&$0.3$ &   $4.653$ & $0.293$ & $0.279$ & $3.235$  & $00.00$\\
$$&$0.5$ &   $4.972$ & $0.310$ & $0.294$ & $3.359$  & $00.00$\\
$$&$0.8$ &   $5.504$ & $0.339$ & $0.320$ & $3.586$  & $00.00$\\
$$&$1.0$ &   $5.888$ & $0.361$ & $0.339$ & $3.775$ & $00.00$ \\

\hline
$$&$0.0$ &  $5.710$ & $0.339$ & $0.320$ & $2.465$ & $-24.90$\\
$$&$0.1 $ &  $5.837$ & $0.346$ & $0.327$ & $2.544$ & $-22.91$\\
$$&$0.2$ &  $5.970$ & $0.354$ & $0.333$ & $2.605$ & $-22.07$\\
$-0.001$&$0.3$ &   $6.103$ & $0.361$ & $0.340$ & $2.682$ & $-20.61$\\
$$&$0.5$ &  $6.374$ & $0.377$ & $0.354$ & $2.859$  & $-17.48$\\
$$&$0.8$ &   $6.811$ & $0.385$ & $0.361$ & $2.764$  & $-29.73$\\
$$&$1.0$ &   $7.169$ & $0.399$ & $0.372$ & $2.711$ & $-39.24$ \\

\hline
$$&$0.0$ &  $4.716$ & $0.246$ & $0.208$ & $3.235$ & $+04.82$\\
$$&$0.1 $ &  $4.853$ & $0.254$ & $0.213$ & $3.302$ & $+05.29$\\
$$&$0.2$ &  $4.992$ & $0.260$ & $0.218$ & $3.372$ & $+05.69$\\
$-2$&$0.3$ &   $5.136$ & $0.267$ & $0.224$ & $3.451$  & $+06.25$\\
$$&$0.5$ &   $5.426$ & $0.282$ & $0.235$ & $3.406$  & $+01.37$\\
$$&$0.8$ &  $5.948$ & $0.291$ & $0.241$ & $3.300$  & $-08.66$\\
$$&$1.0$ &   $6.382$ & $0.302$ & $0.249$ & $3.205$ & $-17.78$ \\

\hline
$$&$0.0$ &   $4.158$ & $0.200$ & $0.144$ & $4.744$ & $+35.09$\\
$$&$0.1 $ &   $4.300$ & $0.207$ & $0.148$ & $4.872$ & $+35.81$\\
$$&$0.2$ &   $4.445$ & $0.213$ & $0.152$ & $5.067$ & $+36.48$\\
$-3$&$0.3$ &   $4.609$ & $0.221$ & $0.157$ & $5.181$  & $+37.56$\\
$$&$0.5$ &   $4.898$ & $0.232$ & $0.163$ & $5.282$  & $+36.40$\\
$$&$0.8$ &  $5.479$ & $0.244$ & $0.170$ & $4.932$  & $+27.29$\\
$$&$1.0$ &   $5.982$ & $0.258$ & $0.178$ & $4.909$ & $+23.10$ \\
\hline
\hline
\end{tabular*}
\end{table*}
\subsection{Effects of Central Magnetic Field $B_{0}$ on Magnetized NS Structure}
This subsection studies the effect of $B_{0}$ on the maximum mass
and radius of magnetized NSs. It can be seen from Fig. \ref{9}
that in the presence of magnetic field, the maximum mass and the
corresponding radius increase.
 As presented in Table \ref{X}, the maximum mass and  corresponding radius  of magnetized NSs are computed for varying values of the central magnetic field $B_{0}$ within the framework of  $f(\mathcal{R},T)$ gravity. Our analysis reveals that an increase in $B_{0}$   leads to an enhancement in the maximum mass or radius. This behavior is further illustrated in Fig. \ref{9}, which demonstrates that the mass-radius relation exhibits a  dependence on $B_{0}$ (for further details see Table \ref{X}).
\begin{table*}
\caption{Effect of central magnetic field $B_{0}$ on magnetized NS structure with $\lambda =-2.0$ and $B_{surf}=5.0\times 10^{16} G$ .}
\label{X}
\begin{tabular*}{\textwidth}{@{\extracolsep{\fill}}lrrrrrrl@{}}
\hline
\hline
$B_{0} [G] $ & \multicolumn{1}{c}{ ${M_{max}}\ [M_{\odot}]$} & \multicolumn{1}{c}{$R\ [km]$} & \multicolumn{1}{c}{$R_{Sch}\ [km]$} & \multicolumn{1}{c}{ $C$} & \multicolumn{1}{c}{$z$} & \multicolumn{1}{c}{$K[10^{-8}$ $m^{-2}]$}  \\
\hline
$0.0$ & $2.345$ & $10.413$ & $5.271$ & $0.253$ & $0.213$ & $2.126$ \\
$1.0 \times 10^{18}$ & $2.780$ & $11.922$ & $6.249$ & $0.262$ & $0.220$ & $1.679$ \\
$2.0 \times 10^{18}$ & $2.710$ & $11.790$ & $6.092$ & $0.258$ & $0.217$ & $1.692$  \\

\hline
\hline
\end{tabular*}
\end{table*}
\section{ Other Structural Parameters of Neutron Stars} \label{Sec.V}
\begin{figure}[h!]
    \includegraphics[width=8.5cm]{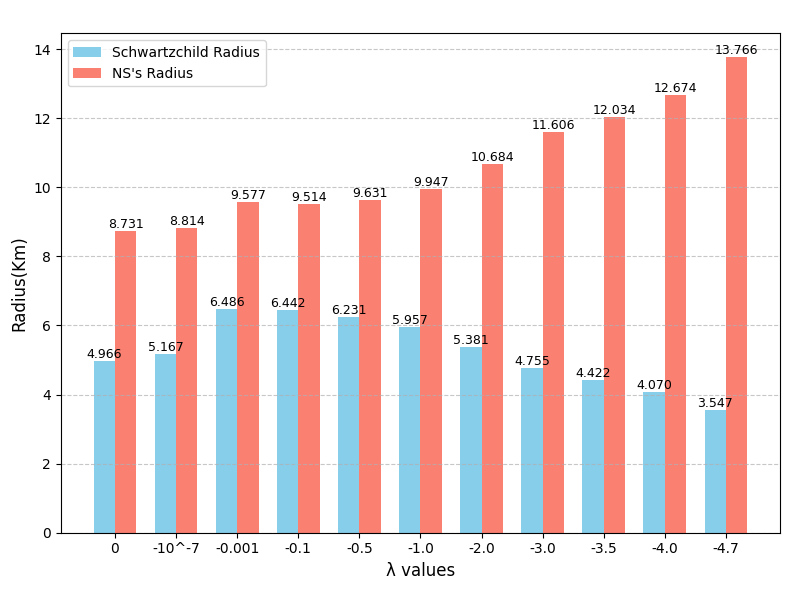}
    \caption{Comparison of the Schwarzschild radius with the stellar radius for different values of the coupling parameter ($\lambda$). The Schwarzschild radius exhibits a strong dependence on the parameter $\lambda$, As illustrated in our analysis. Notably, We observe a non-monotonic relationship; While the radius initially increases with the coupling parameter, $R_{Sch}$ peaks before subsequently decreasing. This behavior suggests a critical transition in the system's gravitational properties, potentially indicative of competing effects between the parameter $\lambda$ and spacetime curvature.}  \label{4}
\end{figure}

In this section, We examine  physical properties of NSs, including
their Schwarzschild radius, compactness, gravitational surface
redshift, and Kretschmann scalar, among others. These properties
provide critical insights into the structural and gravitational
characteristics of NSs, enabling us to distinguish them from other
compact objects such as black holes and white dwarfs
\cite{Karami-46}. These apply to both isotropic and anisotropic
NSs (see Tables  \ref{I} and  \ref{II} and \ref{VI}).
As demonstrated in Table \ref{I}, for a set of various coupling parameters ($\lambda$), we have calculated the aforementioned properties of isotropic, non-magnetized NSs.\\
\textbf{Schwarzschild Radius:} To determine the nature of compact objects in our study, we compute the Schwarzschild radius $R_{Sch}$ for each mass configuration under varying magnetic fields (see Tables  \ref{I} and  \ref{II} and \ref{VI}). The Schwarzschild radius in GR is given by $R_{Sch}^{GR}=\frac{2GM}{c^2}$, while in $f(\mathcal{R},T)$ gravity is as follows,
\begin{equation}
R_{Sch}^{f(\mathcal{R},T)}=\frac{2GM}{c^2}\biggl(1+\frac{3\lambda}{8\pi}\biggr).
\label{41}
\end{equation}
For $B_{surf}=5.0\times 10^{16} G$, our calculations yield a
maximum Schwarzschild radius of $7.573\ km$, confirming that the
compact objects in our system cannot be black holes. Instead, the
observed radius is $10.003\ km$ consistent with NSs, as they
significantly exceed the theoretical $R_{Sch}$ limit. Furthermore,
our analysis reveals a positive correlation between the magnetic
field strength and the Schwarzschild radius (see Tables \ref{II}
and \ref{VI}). This suggests that the strong magnetic fields
influence the effective spacetime geometry in $ f(\mathcal{R},T)$
gravity, leading to an increase in $R_{Sch}$. Fig. \ref{4}  and
Table \ref{I}  illustrate something crucial, the Schwarzchild
radius is strongly sensitive to the coupling parameter
($\lambda$), which with the diminishing of $\lambda$ the
Schwarzchild radius shrinks (see Fig.\ref{4} and Eq. (\ref{41}))
and at ($-\frac{8\pi}{3}$) it completely vanish. These results
provide critical constraints on the viability of modified gravity
theories in describing magnetized NSs.
\\
\textbf{Compactness:} Compactness is a fundamental property of NSs, characterizing the strength of their surface gravitational field. For a spherical object, compactness $(C)$ is defined as the ratio of the mass to the stellar radius $C \equiv \frac{M}{R}$, where $C$ quantifies the influence of strong-field gravity. Our analysis reveals a correlation between increasing magnetic field strength and enhanced compactness, as demonstrated by the numerical results presented in Tables \ref{II} and \ref{VI}. This suggests that the magnetic fields play a non-negligible role in modifying the spacetime geometry around NSs.
\\
\textbf{Surface Redshift:} Another known parameter of NSs is the gravitational surface
redshift. In GR, the surface redshift is $z\equiv \frac{1}{\sqrt{g_{00}}} -1=e^{-\Phi (R)/2} -1$, while in $ f(\mathcal{R},T)$ gravity, $z$ is derived from Eqs. (\ref{19}) and (\ref{20}) as follow,
\begin{equation}
\Phi (R)=\int _{0}^{R} \frac{\frac{Gm(r)}{c^2}+4\pi r^3 \biggl[P(r)+\frac{\lambda }{8\pi +2\lambda}(\epsilon(r) c^2- P(r))\biggr]}{r^2 (1-\frac{2Gm(r)}{rc^2})} dr.
\label{42}
\end{equation}
If we use the boundary conditions $m(R)=M$, $P(R)=0$ and $\epsilon (R)=\epsilon_{c}$, then Eq. (\ref{42}) yields,
\begin{equation}
e^{-\Phi(R)/2} = exp\biggl[\frac{GM}{Rc^2}\biggl(1+\frac{3\lambda }{8\pi +2\lambda}\biggr)\biggl(1-\frac{2GM}{3Rc^2}\biggr)\biggr].
\label{43}
\end{equation}
By expanding the above equation and defining $\Xi \equiv 1+\frac{3\lambda}{8\pi +2\lambda}$, finally we get,
\begin{equation}
z \approx \Xi \frac{GM}{Rc^2}+\biggl(\frac{\Xi^2}{2}-\frac{2 \Xi}{3}\biggr)\frac{G^2M^2}{R^2c^4} .
\label{44}
\end{equation}
Obviously, if  $\lambda$ tends to zero, it will match the standard weak-field gravitational redshift in GR up to second order in $\frac{GM}{Rc^2}$. The values of $z$ are shown in Tables \ref{II} and \ref{VI}.

\noindent \textbf{Kretschmann Scalar:} In the study of any spacetime, a fundamental question is whether the geometry is regular, that is, whether all curvature invariants remain finite at every point. A spacetime is singular if at least one curvature invariant diverges. For the Schwarzschild metric, the Ricci tensor $\mathcal{R}_{\mu\nu}$ and Ricci scalar $\mathcal{R}$ vanish identically in vacuum, providing no insight into the underlying curvature structure. Thus, deeper geometric quantities must be examined. The Riemann curvature tensor $\mathcal{R}_{\mu\nu\alpha\beta}$ encodes the full tidal and geodesic deviation properties of spacetime. However, due to its complexity, scalar invariants derived from it are often employed. The Kretschmann scalar $K^2=\mathcal{R}_{\mu\nu\alpha\beta} \mathcal{R}^{\mu\nu\alpha\beta}$ is particularly useful in vacuum, as it remains non-zero and characterizes curvature without coordinate artifacts. For a NS, the curvature at its surface can be expressed as,
\begin{eqnarray}
&&K^2\approx \frac{48G^2[m^{GR}(R)]^2}{R^6 c^4}+\frac{96G^2m^{GR}(R)m^{f(\mathcal{R},T)}(R)}{R^6 c^4 }\lambda
 \nonumber\\
&&+\frac{48G^2[m^{f(\mathcal{R},T)}(R)]^2}{R^6 c^4}\lambda^2 .\,  \label{45}
\end{eqnarray}
For each value of $\lambda$, our results of $K$ (Table \ref{II} and Table \ref{VI}) reveal a peak.
Notably, we observe that increasing the star's magnetic field amplifies the gravitational curvature, suggesting a coupling between electromagnetism and strong-field gravity.
\begin{table*}
\caption{Comparsion of our results with observational measurements (constraints on our model). Here, ($Th$) and ($Ob$) denotes theoretical and observational findings, respectively. }
\label{XI}
\begin{tabular*}{\textwidth}{@{\extracolsep{\fill}}lrrrrrrl@{}}
\hline
\hline
$ Compact$ $ Objects$ & \multicolumn{1}{c}{ ${M_{max}}\ [M_{\odot}](Ob)$} & \multicolumn{1}{c}{$R\ [km](Ob]$} & \multicolumn{1}{c}{ $B_{surf}[G]$}& \multicolumn{1}{c}{$\lambda$} & \multicolumn{1}{c}{ $\beta$} & \multicolumn{1}{c}{${M_{max}}\ [M_{\odot}](Th)$} & \multicolumn{1}{c}{$R\ [km](Th)$} \\
\hline

PSR J0952-0607\cite{Romani:2022jhd-72}  & $2.35$ & $-$  & $1.0\times 10^{17}$ & $-3$ & $0.2$ & $2.345$ & $10.402$ \\
PSR J0740+6620  \cite{romani12-73} & $2.08^{+0.07}_{-0.08}$ & $13.7^{+2.6}_{-1.5}$  & $5.0\times 10^{16}$ & $-0.001$ & $0.1$ & $2.079$ & $8.714$\\
 PSR J0348+0432  \cite{Miller21-74} & $1.97^{+0.04}_{-0.04}$ & $10\pm2.0$ & $1.0\times 10^{17}$  & $-0.001$ & $0.1$ & $1.977$ & $8.415$ \\
PSR J0952-0607  \cite{Miller21-74} & $2.35\pm 0.17$ & $-$ & $5.0\times 10^{16}$  & $-2 $ & $0.2$ & $2.333$ & $9.942$ \\
 Vela X-1   \cite{Kretschmar:2019hyk-76} & $1.88^{+0.13}_{-0.13}$ & $9.56$  & $1.0\times 10^{17}$ & $0.0$ & $0.8$ & $1.864$ & $8.101$ \\
 4U 1608-52  \cite{Bhattacherjee24-77} & $1.74^{+0.14}_{-0.14}$ & $9.3\pm1.0$  & $5.0\times 10^{16}$ & $0.0$ & $0.5$ & $1.778$ & $8.309$ \\
 PSR J16-2230 \cite{Lenzi:2013dya-78}  & $1.97$ & $12\pm2.0$  & $1.0\times 10^{17}$ & $-0.001$ & $0.1$ & $1.977$ & $8.414$\\
PSR J1311-3430 \cite{sul24-05} & $2.7$ & $-$  & $5.0\times 10^{16}$ & $-3$ & $0.5$ & $2.719$ & $11.366$\\
 PSR J0030+0451 \cite{Miller:2019cac}  & $ 1.43^{+0.20}_{-0.17}$ & $13.02^{+1.24}_{-1.06}$  & $1.0\times 10^{17}$ & $0.0$ & $0.0$ & $1.431$ & $7.805$\\
 GW170817 \cite{abbott19-4}  & $1.46^{+0.12}_{-0.10}$ & $10.8^{+2.0}_{-1.7}$  & $1.0\times 10^{17}$ & $0.0$ & $0.1$ & $1.477$ & $7.847$\\
 GW190425 \cite{abbott20-79} & $1.6-1.87$ & $-$  & $5.0\times 10^{16}$ & $0.0$ & $0.2$ & $1.620$ & $8.206$\\
 GW190814 \cite{LIGOScientific:2020zkf}  & $2.59^{+0.08}_{-0.09}$ & $-$  & $1.0\times 10^{17}$ & $-3$ & $0.5$ & $2.584$ & $10.553$\\
 GW190917 \cite{abbott23-81}  & $2.10^{+1.1}_{-0.4}$ & $-$  & $5.0\times 10^{16}$ & $-0.001$ & $0.2$ & $2.122$ & $8.720$\\
 GW200105 \cite{LIGOScientific:2021qlt-70} & $1.9^{+0.3}_{-0.2}$ & $-$  & $5.0\times 10^{16}$ & $0.0$ & $0.8$ & $1.954$ & $8.385$\\
 PSR J2215+5135  \cite{Sullivan:2024qgl} & $2.27^{+0.17}_{-0.15}$ & $-$ & $5.0\times 10^{16}$  & $-2$ & $0.1$ & $2.272$ & $9.933$ \\
 GW200210-092254 \cite{KAGRA:2021duu, Maurya_2024}  & $2.83^{+0.47}_{-0.42}$ & $-$  & $1.0\times 10^{17}$ & $-2 $ & $1.0$ & $2.839$ & $10.554$\\
 GW230529 A \cite{Janquart:2024ztv-84} & $3.6^{+0.8}_{-1.2}$ & $-$ & $5.0\times 10^{16}$ & $-3 $ & $1.0$ & $3.337$ & $12.282 $\\

\hline
\hline
\end{tabular*}
\end{table*}

\noindent { \textbf{The deviation of Kretschmann scalar in $f(\mathcal{R},T)$
gravity  from that in GR:}  If  $\lambda \to 0  $, then the Kretchmann
scalar tends to its GR form as follows,
\begin{equation}
Q^2 = \frac{48G^2M^2}{R^6 c^4}.
\label{46}
\end{equation}
Here, we compare $K$ and $Q$ by introducing the deviation parameter $d$ as follows,
\begin{equation}
d=\frac{K-Q}{K}\times100.
\label{47}
\end{equation}
The results are presented in Tables \ref{II} and \ref{VI}.  These indicate that for certain values of $\lambda$ and $\beta$, the deviation parameter ($d$) shows a peak corresponding to the peak in $K$.}

\section{Observational Constraints on Our Model} \label{Sec.VI}
Here, We perform a comparative analysis with observational data
and demonstrate the observational constraints imposed by our
model. Hence, we present our observational data and theoretical
results in Table  \ref{XI}. Our analysis concludes that
"reproducing" the observational outcomes necessitates imposing
specific constraints on our computational model. For the
gravitational model used in this work, in addition to the equation
of state, consistency with observational data requires that its
parameters satisfy specific constraints. These constraints are
detailed in Table  \ref{XI}. For instance, to replicate the
results for  PSR J0740+6620  within our anisotropic model in
$f(R,T)$ gravity, for $B_{surf}=1.0\times 10^{17} G$, the free
parameters $\lambda$ and  $\beta$ must be set to $-0.001$ and
$0.3$, respectively. Furthermore, these observational results
impose specific numerical constraints on the parameters of the
$f(\mathcal{R},T)$ gravity theory itself.  In addition to these
data, the two pulsars indicated in all panels of Fig. \ref{12}
also impose the constraints on the parameters of this theory in
Table  \ref{XI}. We set the parameters such that, in addition to
satisfying the mass and radius of GW170817 and the masses of the secondary components  GW190814 and GW200210-092254, the coupling and
anisotropy parameters we employed not only fulfill the constraints
from the massive NICER pulsars (which are also referred to as
massive NSs), but also simultaneously account for those unknown
masses within the mass gap. As detailed in Table  \ref{XI}, to
reproduce the results of an event like the mass of GW230529A:(i) For
$B_{surf}=5.0\times 10^{16} G$, the free parameters $\lambda$ and
$\beta$ must take the values $-2.0$ and $1.0$. (ii) Conversely,
for $B_{surf}=1.0\times 10^{17} G$, the values must be $-3.0$ and
$1.0$, respectively.
\section{Summary and Conclusions} \label{Sec.VII}
In this work, we computed the structure properties of neutron star
(NS) within the framework of $f(\mathcal{R},T)$ gravity for both
isotropic and anisotropic cases. In the structure calculations, we
employed the equation of state (EoS) derived from a microscopic
computation using the accurate nucleon-nucleon interactions, the
AV18 potential. The present analysis is based on a single representative EoS (i.e. AV18), a specific anisotropy prescription (i.e. Bowers-Liang model), and a particular functional form of 
$f(\mathcal{R},T)$ modified gravity. Hence, the results are model-dependent and should be interpreted accordingly.
\par In our calculations, we utilized a Gaussian form
for the magnetic field. For this purpose, at first using the EoS
for an isotropic non-magnetized NS, we solved the modified TOV
equations of $f(\mathcal{R},T)$ gravity. We then incorporated the
Bowers-Liang anisotropic model, and applied it to an anisotropic
magnetized NS, solving the corresponding TOV equations in
$f(\mathcal{R},T)$ gravity to study its structure. In the
isotropic case, we found that a decreasing in coupling parameter
yields higher masses and radii. For instance, $\lambda=-3$
resulted in a mass of $2.509\ M_{\odot}$. Consequently, in the
anisotropic case, by varying $\lambda $ and $\beta$, we were able
to predict even higher masses and radii. A configuration with
$\beta=1$ and $\lambda=-3$ yielded a mass of $3.337\ M_{\odot}$,
which is highly suitable for predicting the objects within the
mass gap. For the anisotropic magnetized NS, we assumed two
surface magnetic field strengths: one at $5.0 \times 10^{16}\ G$
and the other one at $1.0 \times 10^{17}\ G$. Since the EoS
becomes softer under the stronger second magnetic field, it
produces lower maximum masses and corresponding radii compared to
the first one. However, it is a well-established and
observationally noted phenomenon that increasing the star's
overall magnetic field (i.e., the central plus surface magnetic
fields) leads to an increase in mass. Throughout our analysis, for
a fixed $\lambda $, an increase in the parameter $\beta$ leads to
an increase in the corresponding masses and radii. Conversely, for
a fixed $\beta$, a decrease in the coupling parameter ($\lambda $)
results in larger masses and radii. Furthermore, we calculated
some physical parameters of the NS such as the Schwarzschild
radius, compactness, surface gravitational redshift, and
Kretschmann scalar in this model. 
Our results demonstrate that the
objects found in the mass gap in a linear model of  $f(\mathcal{R},T)=\mathcal{R}+2 \lambda T$ gravity might be interpreted as NSs. 
In addition, through the computation of the Schwarzschild radius and the compactness, 
we have shown that the compact objects resulting from our analysis  cannot be identified as black holes.
 This conclusion is supported by the fact that their Schwarzschild radii
are all smaller than the stellar radius (i.e., compactness is less
than one), and the surface gravitational redshift is below $0.4$,
which is typical for NSs. Crucially, the Kretschmann scalar yields
finite values for all computed models. 
{0.00,0.07,1.00}{Collectively,  in  $f(\mathcal{R},T)$ gravity, these physical parameters confirm that the objects observed within the
mass gap might be interpreted as neutron stars.}
%
Finally, a comparison is made between our results and the observational data from LIGO/Virgo/KAGRA and NICER. The parameters of the $f(\mathcal{R},T)$ theory, along with the anisotropy parameter, are set in a manner that successfully reproduces the masses and radii of GW170817, PSR J0952-0607, and PSR J0740+6620. For the secondary components of GW190814 and GW200210-092254 there is no information about the radii, however, their masses are successfully reproduced in this work.
It is important to emphasize that the present analysis is based on a specific EoS
(i.e. AV18), a specific anisotropy prescription (i.e. Bowers - Liang), and a
particular functional form of $f(\mathcal{R},T)$ modified gravity. On the other hand, it is
well known that the results of compact objects structure calculations are model dependent. In fact a fully robust constraint on the nature of compact objects in
the mass gap region would require a systematic treatment of EoS uncertainties,
for instance, through marginalization over a wide class of EoS models, as well as a
joint statistical inference of both gravitational and matter-sector parameters.
However, in this work we have employed an accurate EoS can actually lead to
more reliable outcomes.
\section*{Acknowledgements}
We wish to thank Shiraz University Research Council.


\end{document}